\documentclass[aps,twocolumn,superscriptaddress,showpacs,pra,amssymb, amsmath,latexsym]{revtex4}
\newtheorem{thm}    {Theorem}
\newtheorem{lem}     {Lemma}
\newtheorem{cor}  {Corollary}
\newtheorem{rem}     {Remark}
\def\QED{\mbox{\rule[0pt]{1.5ex}{1.5ex}}}

\def\endproof{\hspace*{\fill}~\QED\par\endtrivlist\unskip}
\def\im{\mathop{\bf Im}\nolimits}
\def\re{\mathop{\bf Re}\nolimits}
\def\tr{\mathop{\rm tr}\nolimits}
\def\Tr{\mathop{\rm Tr}\nolimits}

\def\real{\mathbb{R}}

\def\complex{\mathbb{C}}

\def\SU{\mathop{\rm SU}\nolimits}
\def\su{\mathop{\mathfrak{su}}\nolimits}

\newcommand{\defeq}{\stackrel{\rm def}{=}}

\def\Label#1{\label{#1}\ [\ #1\ ]\ }
\def\Label{\label}

\begin{document}

\title{Asymptotic performance of optimal state estimation in quantum two
level system}
\author{Masahito Hayashi}
\email{masahito@qci.jst.go.jp}
\address{Quantum Computation and Information Project, ERATO-SORST, JST\\
5-28-3, Hongo, Bunkyo-ku, Tokyo, 113-0033, Japan}
\author{Keiji Matsumoto$^{\rm 1}$}
\email{keiji@nii.ac.jp}
\address{National Institute of Informatics,
2-1-2 Hitotsubashi, Chiyoda-ku, Tokyo 101-8430,Japan}
\pacs{03.65.Wj,03.65.Ud,02.20.-a}
\begin{abstract}
We derived an asymptotic bound the accuracy of the estimation when 
we use the quantum correlation in the measuring apparatus.
It is also proved that 
this bound can be achieved in any model in the quantum two-level system.
Moreover, we show that this bound of such a model cannot be attained by 
any quantum measurement with no quantum correlation 
in the measuring apparatus.
That is, in such a model, the quantum correlation can improve 
the accuracy of the estimation in an asymptotic setting.
\end{abstract}
\maketitle

\section{Introduction}
Estimating unknown quantum state is an important task in quantum information.
In this paper, we discuss this problem by focusing on 
two typical quantum effects;
One is the uncertainty caused by the non-commutativity.
The other is the quantum correlation between particles,
{\em e.g.}, quantum interference, quantum entanglement, {\em etc}.
Indeed, the probabilistic property in quantum mechanics
is caused by the first effect.
Hence, it is impossible to determine the initial quantum state
based only on the single measurement.
Due to this property, we need some statistical processing
for identifying the unknown state.
Needless to say, it is appropriate for effective processing
to use a measurement drawing much information.
Therefore, the optimization of measuring process is
important for this purpose.
The second property is also crucial for this optimization.
This is because it is possible to use the quantum correlation between 
several particles.
Hence, we compare the optimal performance 
in presence or absence of quantum correlation between 
several particles in the measuring process.
This paper treat this comparison in the case of two-dimensional case,
i.e., the qubit case.

Estimating unknown quantum state is 
often formulated as the identification problem
of the unknown state which is assumed to be included
a certain parametric quantum states family.
Such a problem is called quantum estimation,
and has been one of the main issues of quantum statistical inference.
In this case, we often adopt mean square error (MSE) as
our error criterion.
As is known in mathematical statistics,
the MSE of an estimator is almost proportional to the inverse of
the number $n$ of observations.
Hence, concerning the estimation of quantum state,
that of an estimator is almost proportional to the inverse of
the number $n$ of systems prepared identically.
Especially, it is the central issue to calculate
the optimal coefficient of it.
Moreover, 
for this purpose, 
as is discussed by Nagaoka\cite{Na2}, Hayashi \& Matsumoto\cite{HM}, 
Gill \& Massar\cite{GM},
it is sufficient to minimize the MSE at a local setting.
(For detail, see section \ref{25-1}.)

The research of quantum estimation has been initiated by
Helstrom\cite{Hel1}.
He generally solved this problem 
in the one-parameter case at the local setting.
However, the multi-parameter case is more difficult
because we need to treat the trade-off among
the MSEs of the respective parameters.
That is, we cannot simultaneously realize 
the optimal estimation of the respective parameters.
This difficulty is caused by the non-commutativity.
First, Yuen \& Lax \cite{YL} and Holevo \cite{HolP} derived 
the bound of the estimation performance 
in the estimation of quantum Gaussian family.
In order to treat this trade-off,
they minimized the sum of the weighted sum of 
the MSEs of respective parameters.
Especially, Yuen \& Lax treated the equivalent sum,
and Holevo did the generally weighted sum.

After this achievement,
Nagaoka \cite{Na}, Fujiwara \& Nagaoka\cite{FN}, Hayashi\cite{Ha}, 
Gill \& Massar \cite{GM}
calculated that of the estimation in the quantum two level system.
They also adopt the same criterion.
Concerning the pure states case,
Fujiwara \& Nagaoka\cite{FN2}, and Matsumoto\cite{Mat} proceeded to 
more precise treatments.


However, the above papers did not treat 
the performance bound of estimation with 
quantum correlation in measuring apparatus,
which is one of the most important quantum effects.
In this paper, we discuss
whether the quantum correlation can improve 
its performance.
For this purpose, we calculate
the CR bound, {\em i.e.},
the optimal decreasing coefficient of the sum of MSEs
with quantum correlation in measuring apparatus,
in the several specific model.


First, as a preparation, we focus on quantum Gaussian family,
and prove that the above quantum correlation 
has no advantage for estimating the unknown state
at section \ref{Gau}.
The reason is roughly given by the following two facts.
One is the fact that the optimal error without quantum correlation
is given by the right logarithmic derivative (RLD) Fisher Information matrix,
which is one of quantum analogues of Fisher Information matrix.
The second is the fact that
the CR bound can be bounded by RLD Fisher Information matrix.

Next, we proceed to the quantum two-level system,
which can be regarded as the quantum analogue of 
the binomial distribution.
In this case, as is shown in section \ref{27-15},
quantum quantum correlation can improve
the performance of estimation.
As the first step, we focus on 
the equivalent sum of the MSEs of respective parameters 
in the parameterization
$\frac{1}{2}
\left(
\begin{array}{cc}
1+ z & x+i y \\
x- iy & 1- z\\
\end{array}
\right)$
with the parameter $x^2 +y^2 +z^2  \le 1$.
As is discussed in subsection \ref{Cov},
the asymptotically optimal estimator is given as follows.

When the quantum state is parameterized in another way:
$
\frac{1}{2}\left(
\begin{array}{cc}
1+ r \cos 2 \theta & r e^{i \phi} \sin 2 \theta \\
r e^{-i \phi} \sin 2 \theta & 1 - r \cos 2 \theta \\
\end{array}\right)$
with the parameter $0 \le r \le 1, 0 \le \phi\le 2\pi , 0 \le \theta\le 
\frac{\pi}{2}$,
we can divide our estimation into two parts.
One is the estimation of $r$,
the other is that of the angle $(\theta,\phi)$.

The estimation of $r$ can be realized by performing 
the projection measurement corresponding to 
the irreducible decomposition of the tensor product representation of
$\SU(2)$,
which equals the special case of 
the measurement used in Keyl \& Werner\cite{KW}, 
Hayashi \& Matsumoto\cite{HM2}.
Note that they derived its error with the large deviation criterion,
but did not treat its MSE.
After this measurement, we perform a covariant measurement for
the estimation of $(\theta,\phi)$.
By calculating the asymptotic behavior of the sum of its MSEs
of respective parameters,
it can be checked that
it attains its lower bound given by RLD Fisher information,
asymptotically.
That is, this estimator is shown to be the optimal with
the above mentioned criterion.
Finally, by comparing the optimal coefficient without
quantum correlation in measuring apparatus,
we check that using this quantum effect can improve
the estimation error.
Furthermore, we treat the CR bound with the general weight matrix
by a more technical method in subsection \ref{Gen}.
In this discussion, the key point is
the fact that 
this model can be asymptotically
approximated by quantum Gaussian model.

This paper is organized as follows.
First, we discuss the lower bounds of asymptotic error
in section \ref{25-1},
which contains reviews of the previous results.
In section \ref{Gau}, quantum Gaussian model is discussed.
We discuss the asymptotic approximation of spin $j$ system by 
the quantum Gaussian model in section \ref{27-11}.
Using these preliminaries, we treat 
quantum two level system in section \ref{27-15}.

\section{Lower bounds of estimation error}\Label{25-1}
\subsection{Quasi Quantum CR bound}
Let ${\it \Theta}$ be an open set in $\real^d$, and let ${\cal S}
=\{ \rho_\theta ;\theta \in {\it \Theta} \}$ be a family of density 
operators on a Hilbert space ${\cal H}$ smoothly parameterized by a
$d$-dimensional parameter $\theta = (\theta^1,\ldots\,,\theta^d)$ 
with the range ${\it \Theta}$.  Such a family is called an $d$-dimensional 
quantum statistical model.  We consider the parameter estimation problem 
for the model ${\cal S}$, and, for simplicity, assume that
any element $\rho_\theta$ is strictly positive.
The purpose of the theory is to 
obtain the best estimator and its accuracy.
The optimization is done by the appropriate choice of
the measuring apparatus and the function from
data to the estimate.

Let $\sigma(\Omega)$ be a $\sigma$- field in the space $\Omega$.
Whatever apparatus is used,
the data $\omega\in\Omega$
lie in a measurable subset $B\in \sigma(\Omega)$
of $\Omega$ writes
\begin{eqnarray*}
{\rm Pr}\{ \omega \in B|\theta \} =
{\rm P}^M_\theta(B)\defeq
\Tr \rho_{\theta}M(B),
\end{eqnarray*}
when the true value of the parameter is $\theta$.
Here, $M$, which is called positive operator-valued measure
(POVM, in short),
is a mapping  
from subsets $B\subset \Omega$ to non-negative Hermitian operators in 
${\cal H}$, such that
\begin{align*}
&M(\emptyset)=O,\, M(\Omega)=I \\
& M(\bigcup_{j=1}^{\infty} B_j),
\!=\!\sum_{j=1}^{\infty}M(B_j)
\quad(B_k\cap B_j=\emptyset,k\neq j)
\end{align*}
(see p.~53 \cite{Hel} 
and p.~50 \cite{HolP}).
Conversely, some apparatus corresponds to any POVM $M$.
Therefore, we refer to the measurement which is controlled by
the POVM $M$ as `measurement $M$'.
Moreover, for estimating the unknown parameter $\theta$, we need an
estimating function $\hat{\theta}$ mapping the observed data $\omega$
to the parameter.
Then, a pair $(\hat\theta,M)$
is called an estimator. 

In estimation theory, we often focus on the unbiasedness condition:
\begin{align}
\int_{\Omega}
\hat{\theta}^j(\omega) \Tr M(\,d \omega) \rho_\theta
= \theta^j , \quad\forall \theta \in\Theta.\Label{ubs1}
\end{align}
Differentiating this equation, we obtain
\begin{align}
\int_{\Omega}
\,\hat{\theta}^j(\omega) \,\frac{\partial}{\partial \theta^k} 
\Tr M(\,d \omega)\rho_\theta
\,=\, \delta^j_k \quad (j,k\,=\,1,2,\ldots ,n), \Label{ubs2}
\end{align}
where $\delta^j_k$ is the Kronecker's delta.  
When $(\hat\theta,M)$ satisfies (\ref{ubs1}) and (\ref{ubs2}) 
at a fixed point $\theta\in{\it \Theta}$, we say that 
$(\hat\theta,M)$ is locally unbiased at $\theta$.  Obviously, 
an estimator is unbiased if and only if it is locally 
unbiased at every $\theta\in{\it \Theta}$. 
In this notation, we 
often describe the accuracy of the estimation at $\theta$ by
the MSE matrix:
\begin{align*}
{\rm V}_\theta^{k,j}
(\hat{\theta},M) \defeq 
\int_{\Omega}
(\hat{\theta}^k -\theta^k )
(\hat{\theta}^j -\theta^j )
\Tr M(\,d \omega) \rho_\theta.
\end{align*}
or 
\begin{align*}
\tr {\rm V}_\theta (\hat\theta,M)G
\end{align*}
for a given weight matrix, which is a positive-definite 
real symmetric matrix. 
Indeed, in the quantum setting, there is not necessarily 
minimum MSE matrix,
while the minimum MSE matrix exists in the classical asymptotic setting.
Thus, we usually focus on $\tr {\rm V}_\theta(\hat\theta,M)$
for a given weight matrix.

We define classical Fisher information matrix $J^M_\theta$ by the POVM $M$
as in classical estimation theory:
\begin{eqnarray*}
J^M_\theta:=
\left[\int_{\omega\in\Omega} 
\partial_i\log\frac{{\rm d}{\rm P}^M_\theta}{{\rm d}\omega}
\partial_j\log\frac{{\rm d}{\rm P}^M_\theta}{{\rm d}\omega}
\,d \omega\right],
\end{eqnarray*}
where $\partial_i=\partial/\partial\theta^i$.
Then, $J^M_\theta$ is characterized, from knowledge of classical 
statistics, by,
\begin{align}
(J^{M}_\theta)^{-1}=\inf_{\hat\theta}
\{
V_\theta(\hat\theta,\,M)\,|\,
\mbox{$(\hat\theta,\,M)$ is locally unbiased}
\},
\Label{14-1}
\end{align}
and the quasi-quantum Cram\'{e}r-Rao type bound
(quasi-quantum CR bound)
$\hat{C}_\theta(G)$ is defined by,
\begin{eqnarray*}
\hat{C}_\theta(G)\defeq \inf\{
\tr G{\rm V}_{\theta}(\hat\theta,\,M)
\:|\:
(\hat\theta,M) \mbox{is locally unbiased}\},
\end{eqnarray*}
and has other expressions.
\begin{align}
&\hat{C}_\theta(G)\nonumber \\
=&
\inf\{
\tr G{\rm V}_{\theta}(\hat\theta,M)
\:|\:
(\hat\theta,M) \mbox{ satisfies the condition (\ref{ubs2})}\}
\Label{27-1}\\
=&
\inf\{\tr G (J^{M}_\theta)^{-1}\:|\: 
M
\mbox{ is a POVM on } {\cal H}\}.\Label{27-2}
\end{align}
As is precisely mentioned latter,
the bound $\hat{C}_\theta(G)$ is uniformally attained by
an adaptive measurement, asymptotically\cite{HM,GM}.
Therefore, $\hat{C}_\theta(G)$ expresses
the bound of the accuracy of the estimation 
without quantum correlation in measurement apparatus.

\subsection{Lower bounds of quasi quantum CR bound}
\subsubsection{SLD bound and RLD bound}
In this subsection,
we treat lower bounds of $\hat{C}_\theta(G)$.
Most easy method for deriving lower bound is 
using quantum analogues Fisher Information matrix.
However, there are two analogues at least, and 
each of them has advantages and disadvantages.
Hence, we need to treat both.
One analogue is symmetric logarithmic derivative
(SLD) Fisher information matrix $J_{\theta;j,k}$:
\begin{align*}
J_{\theta;j,k}\defeq 
\langle L_{\theta;j},L_{\theta;k}\rangle_\theta,
\end{align*}
where
\begin{align*}
\frac{\partial \rho_\theta}{\partial \theta^j}& = 
\rho_\theta \circ L_{\theta;j}\\
\langle X,Y\rangle_\theta &\defeq
\Tr \rho_\theta  (X^*\circ Y) = 
\Tr (\rho_\theta \circ Y) X^* \\
X\circ Y &\defeq \frac{1}{2}(XY+YX),
\end{align*}
and $L_{\theta,j}$ is called its symmetric logarithmic derivative
(SLD).
Another quantum analogue is the right logarithmic derivative
(RLD) Fisher information matrix $\tilde{J}_{\theta;j,k}$:
\begin{align*}
\tilde{J}_{\theta;j,k}\defeq 
\Tr \rho_\theta \tilde{L}_{\theta;k} (\tilde{L}_{\theta;j})^*
= ( \tilde{L}_{\theta;j}, \tilde{L}_{\theta;k})_\theta
\end{align*}
where
\begin{align*}
\frac{\partial \rho_\theta}{\partial \theta^j} = 
\rho_\theta \tilde{L}_{\theta;j},\quad
(A,B)_\theta  \defeq \Tr \rho_\theta B A^*,
\end{align*}
and $\tilde{L}_{\theta,j}$ is called its right logarithmic derivative
(RLD).
\begin{thm}\Label{thm-1}Helstrom\cite{Hel}Holevo\cite{HolP}
If a vector $\vec{X}=[X^1, \ldots, X^d]$ of Hermite matrixes
satisfies the condition:
\begin{align}
\Tr \frac{\partial \rho_\theta}{\partial \theta^k}
X^j = \delta_k^j,\Label{8-27-1}
\end{align}
the matrix $Z_\theta(\vec{X}) $:
\begin{align*}
Z_\theta^{k,j} (\vec{X}) \defeq \Tr \rho_\theta X^k X^j
\end{align*}
satisfies the inequalities
\begin{align}
Z_\theta (\vec{X}) \ge (J_\theta)^{-1} \Label{sld1}
\end{align}
and
\begin{align}
Z_\theta (\vec{X}) \ge (\tilde{J}_\theta)^{-1}.\Label{rld1}
\end{align}
\end{thm}
For a proof, see Appendix \ref{10-4-1}.
Moreover, the following lemma is known.
\begin{lem}\Label{le-B}Holevo\cite{HolP}
When we define th vector of Hermitian matrixes
$\vec{X}_M$:
\begin{align*}
X^j_M = \int_{\real^d}
(\hat{\theta}^j - \theta^j)M(\,d \hat\theta) ,
\end{align*}
then
\begin{align}
V_\theta(M) \ge Z_\theta (\vec{X}_M) \Label{29-3}.
\end{align}
\end{lem}
For a proof, see Appendix \ref{10-4-2}.
Combining Theorem \ref{thm-1} and Lemma \ref{le-B},
we obtain the following corollary.
\begin{cor}
If an estimator $M$ is
locally unbiased at $\theta\in{\it \Theta}$,
the SLD Cram\'{e}r-Rao inequality
\begin{align}
V_\theta(M) \ge (J_\theta)^{-1}\Label{sld4}
\end{align}
and the RLD Cram\'{e}r-Rao inequality
\begin{align}
V_\theta(M) \ge (\tilde{J}_\theta)^{-1}\Label{rld5}
\end{align}
hold, where,
for simplicity, we regard a POVM $\tilde{M}$ with the out come in $\real^d$
as an estimator in the correspondence
$\tilde{M}= M \circ \hat{\theta}^{-1}$.
\end{cor}
Therefore,
we can easily obtain the inequality
\begin{align*}
\tr V_\theta(M) G \ge \tr (J_\theta)^{-1} G
\end{align*}
when $M$ is locally unbiased at $\theta$.
That is, we obtain the SLD bound:
\begin{align}
C^S_\theta(G) \defeq \tr (J_\theta)^{-1} G \le 
\hat{C}_\theta(G).\Label{27-4}
\end{align}
As was shown by Helstrom\cite{Hel},
the equality (\ref{27-4}) holds for one-parameter case.
However, we need the following lemma for obtaining 
a bound of $\hat{C}_\theta(G)$ from
the RLD Cram\'{e}r-Rao inequality.
\begin{lem}\Label{le8-26}
When a real symmetric matrix $V$ 
and Hermite matrix W satisfy
\begin{align*}
V \ge W,
\end{align*}
then
\begin{align*}
\tr V \ge \tr \re W + \tr | \im W |,
\end{align*}
where $\re W$ ($\im W$) denotes the real part of $W$ 
(the imaginary part of $W$), respectively.
\end{lem}
For a proof, see Appendix \ref{10-4-3}.
Since the RLD Cram\'{e}r-Rao inequality (\ref{rld5}) yields that
any locally unbiased estimator $M$ satisfies
\begin{align*}
\sqrt{G} V_\theta(M) \sqrt{G}\ge 
\sqrt{G}(\tilde{J}_\theta)^{-1}\sqrt{G},
\end{align*}
lemma \ref{le8-26} guarantees that
\begin{align}
&\tr V_\theta(M) G \nonumber \\
\ge &
\tr \sqrt{G} \re (\tilde{J}_\theta)^{-1}\sqrt{G}
+ \tr | \sqrt{G}\im(\tilde{J}_\theta)^{-1}\sqrt{G}|.\Label{rld2}
\end{align}
Thus, we obtain
the RLD bound:
\begin{align}
C^R_\theta(G) &\defeq \tr \sqrt{G} \re (\tilde{J}_\theta)^{-1}\sqrt{G}
+ \tr |\sqrt{G}\im (\tilde{J}_\theta)^{-1}\sqrt{G}|\nonumber\\
& \le 
\hat{C}_\theta(G). \Label{29-9}
\end{align}
For characterizing the relation between
the RLD bound $C^R(G)$ and the SLD bound $C^S(G)$,
we introduce 
the superoperator ${\cal D}_\theta$ 
as follows\cite{HolP}:
\begin{align*}
\rho_\theta \circ {\cal D}_\theta (X) = i [X, \rho].
\end{align*}
This superoperator is called $D$-operator,
and has the following relation with the RLD bound.
\begin{thm}\Label{thm9-25}Holevo\cite{HolP}
When the linear space $T_\theta$ spanned by $
L_{\theta,1},
\ldots, L_{\theta,d}$ is invariant for the action of 
the superoperator ${\cal D}_\theta$,
the inverse of the RLD Fisher information matrix 
is described as
\begin{align}
\tilde{J}_\theta^{-1}=J_\theta^{-1}+\frac{i}{2} J_\theta^{-1} D_\theta J_\theta^{-1},
\Label{29-6}
\end{align}
where the antisymmetric matrix $D_\theta$ is defined by 
\begin{align}
D_{\theta;k,j}\defeq
\langle {\cal D}_\theta (L_{\theta,j}), L_{\theta;k}\rangle_\theta
= i \Tr \rho_\theta [L_{\theta,k},L_{\theta,j}] .\Label{29-7}
\end{align}
Thus, the RLD bound is calculated as
\begin{align}
C^R_\theta(G) = \tr G J_\theta^{-1}
+ \frac{1}{2}\tr |\sqrt{G} J_\theta^{-1} D_\theta J_\theta^{-1}
\sqrt{G}|.
\Label{27-12}
\end{align}
Therefore,
$C^R_\theta(G) \ge C^S_\theta(G) $, {\em i.e.,}
the RLD bound is better than the SLD bound.
\end{thm}
For a proof, see Appendix \ref{10-4-4}.
In the following, we call the model {\em D-invariant},
if the linear space $T_\theta$ is 
invariant for the action of 
the superoperator ${\cal D}_\theta$ for any parameter $\theta$.

\subsubsection{Holevo bound}
Next, we proceed to the non-D-invariant case.
in this case, 
Lemma \ref{le-B} guarantees that 
any locally unbiased estimator $M$ satisfies 
\begin{align*}
\sqrt{G} {\rm V}_\theta(M) \sqrt{G} \ge 
\sqrt{G} Z_\theta(\vec{X}_M) \sqrt{G} ,
\end{align*}
where
\begin{align*}
Z_\theta^{k,j} (\vec{X}) \defeq \Tr \rho_\theta X^k X^j.
\end{align*}
Thus, from Lemma \ref{le8-26},
we have
\begin{align}
&\tr \sqrt{G} {\rm V}_\theta(M) \sqrt{G} \nonumber\\
\ge & C_\theta(G,\vec{X}_M) \nonumber \\
\defeq &\tr  \sqrt{G} \re Z_\theta (\vec{X}_M) \sqrt{G}
 + \tr | \sqrt{G} \im Z_\theta (\vec{X}_M)\sqrt{G}|.
\Label{23-5}
\end{align}
Since $X_M$ satisfies the condition (\ref{8-27-1}),
the relation (\ref{27-1}) yields the following theorem.
\begin{thm}Holevo\cite{HolP}:
The inequality
\begin{align*}
C^H_\theta(G) 
\defeq 
\min_{X}
\left\{
C_\theta(G,\vec{X}) 
\left|
\Tr \frac{\partial \rho_\theta}{\partial \theta^i}
X^j = \delta_i^j
\right. \right\}
\le \hat{C}_\theta(G)
\end{align*}
holds.
\end{thm}
Hence, the bound $C^H_\theta(G)$ is called the Holevo bound.
When $X$ satisfies the condition (\ref{8-27-1}),
the relation (\ref{sld1}) yields that
\begin{align*}
\tr G \re Z_\theta (\vec{X}) =
\tr G Z_\theta (\vec{X}) 
\ge \tr G J_{\theta}^{-1} = C_\theta^S(G),
\end{align*}
which implies
\begin{align*}
C_\theta^H(G) \ge C_\theta^S(G).
\end{align*}
Also, the relation (\ref{rld1}) guarantees that
\begin{align*}
& \sqrt{G} Z_\theta (\vec{X}) \sqrt{G} + 
\left|\sqrt{G} \im Z_\theta (\vec{X})\sqrt{G}\right|\\
\ge &
\sqrt{G} Z_\theta (\vec{X}) \sqrt{G} + 
\sqrt{G} \im Z_\theta (\vec{X})\sqrt{G}
\ge \sqrt{G}
\tilde{J}_\theta\sqrt{G}.
\end{align*}
Similarly to (\ref{rld2}), 
the relation (\ref{rld1}) yields
\begin{align*}
C_\theta(G,\vec{X}) \ge C_\theta^R(G),
\end{align*}
which implies
\begin{align}
C_\theta^H(G) \ge C_\theta^R(G). \Label{rld4}
\end{align}
Moreover, 
the Holevo bound has another characterization.
\begin{lem}\Label{le-3}
Let $\overline{T}_\theta$ be the linear space spanned by 
the orbit of $T_\theta$ with respect to the action of 
${\cal D}_\theta$.
Then, the Holevo bound can be simplified as
\begin{align}
C_\theta^H(G)= 
\min_{\vec{X}:X_k \in \overline{T}_\theta}
\left\{
C_\theta(G,\vec{X}) 
\left|
\Tr \frac{\partial \rho_\theta}{\partial \theta^k}
X^j = \delta_k^j
\right. \right\}.\Label{8-27-2}
\end{align}
Moreover, we assume that
the D-invariant model containing the original model
has normally orthogonal basis $\langle L_1, \ldots, L_m\rangle$
concerning SLD,
and the inverse of its RLD Fisher information matrix
is given by $J$ in this basis.
Then, 
the Holevo bound has the following expression.
\begin{align}
C^H_\theta(G)
= \min_{v=[v^j]}\left\{\left.
\tr | \sqrt{G} 
Z_J(v)
\sqrt{G} |
\right| \re \langle d_k|J|v^j\rangle = \delta_k^j\right\}
\Label{10-19-1}
\end{align}
where
$Z_J^{k,j}(v)\defeq  \langle v^k|J|v^j\rangle$
and 
a vector $d_k$ is chosen as
\begin{align}
\frac{\partial \rho_\theta}{\partial \theta^k}
= \sum_j d_{k,j} \rho \circ L_j.
\end{align}
Note that the vector $v^j$ is a real vector.
\end{lem}
For a proof, see Appendix \ref{10-4-5}.

In the D-invariant case,
only the vector $\vec{L}= [L_{\theta}^k\defeq 
\sum_{j=1}^d (J_\theta^{-1})^{k,j}L_{\theta;j}]$
satisfies the condition in the right hand side (R.H.S.) of (\ref{8-27-2}),
{\em i.e.},
$C_\theta^H(G)= 
C_\theta(G,\vec{L})$.
Since
$\Tr \rho_\theta L_{\theta}^k L_{\theta}^j
= \Tr (\rho_\theta \circ L_{\theta}^k
+ \frac{i}{2}[L_{\theta}^k,\rho_\theta]) L_{\theta}^j$,
the equation (\ref{29-6}) guarantees 
\begin{align}
Z_\theta(\vec{L})= \tilde{J}_\theta^{-1}.
\Label{14-3}
\end{align}
That is, the equation 
\begin{align*}
C_\theta(G,\vec{L})= 
C_\theta^R(G)
\end{align*}
holds.
Therefore, the equality of (\ref{rld4}) holds.

Concerning the non-D-invariant model,
we have the following characterization.
\begin{thm}\Label{thm29-2}
Let ${\cal S}_1
\defeq \{\rho_{(\theta_1, \ldots, \theta_{d_1}, 0, \ldots, 0)}|
(\theta_1, \ldots, \theta_{d_1}) \subset \Theta_1\}
\subset {\cal S}_2
\defeq 
\{\rho_{\theta_1, \ldots, \theta_{d_2}}|
(\theta_1, \ldots, \theta_{d_2}) \subset \Theta_2\}$ 
be two models such that
${\cal S}_2$ is D-invariant.
If a vector of Hermitian matrixes $\vec{X}=[X^k]$ satisfies 
the condition (\ref{8-27-1}) and
$X^k \in <L_{\theta;1},\ldots,L_{\theta;d_2}>$,
then
\begin{align}
C_{\theta,1}(G,\vec{X}) = 
C_{\theta,2}^R(P_{\vec{X}}^T G P_{\vec{X}})
\end{align}
for any weight matrix $G$,
where the $d_1 \times d_2$ matrix $P_{\vec{X}}$ is defined as
\begin{align}
P_{\vec{X};l}^k\defeq
\Tr \frac{\partial \rho_\theta}{\partial \theta^l}X^k ,
\end{align}
{\em i.e.,} $P_{\vec{X}}$ is a linear map from a $d_2$ dimensional space to
a $d_1$ dimensional space.
Furthermore, 
if the bound $C_{\theta,2}^R(P_{\vec{X}} G P_{\vec{X}})$
is attained in the model ${\cal S}_2$, the quantity $C_{\theta,1}
(G,\vec{X})$ can be attained in the model ${\cal S}_2$.
\end{thm}

Here, we denote the linear space spanned by elements $v_1, \ldots, v_l$ by
$<v_1, \ldots, v_l>$.
For a proof, see Appendix \ref{10-4-6}.
Thus, if the RLD bound can be attained for any weight matrix in 
a larger D-invariant model,
the Holevo bound can be attained for any 
weight matrix.

\subsubsection{Optimal MSE matrix and Optimal Fisher information matrix}
\Label{9-29-8}
Next, we characterize POVMs attaining the Holevo bound.
First, we focus on the inequality (\ref{23-5}) 
for a strictly positive matrix $G$.
if and only if 
\begin{align*}
{\rm V}_\theta(M)= 
\re Z_\theta(\vec{X}_M) +
\sqrt{G}^{-1} | \sqrt{G} \im Z_\theta(\vec{X}_M)\sqrt{G}|\sqrt{G}^{-1},
\end{align*} 
the equality of (\ref{23-5}) holds.
Thus, the Holevo bound $C_\theta(G)$ 
is attained for a strictly positive matrix $G$,
if and only if
\begin{align}
{\rm V}_\theta(M)= 
\re Z_\theta(\vec{X}_G) +
\sqrt{G}^{-1} | \sqrt{G} \im Z_\theta(\vec{X}_G)\sqrt{G}|\sqrt{G}^{-1},
\Label{27-9}
\end{align} 
where $\vec{X}_G$ is a vector of Hermitian matrix satisfying 
$C_\theta(G)= C_\theta(G,\vec{X}_G)$.
Therefore, the equation (\ref{14-1}) guarantees that
if and only if the Fisher information matrix 
$J^M_\theta$ of POVM $M$ equals
\begin{align}
\sqrt{G}\left(
\sqrt{G} \re Z_\theta(\vec{X}_G) \sqrt{G}+
| \sqrt{G} \im Z_\theta(\vec{X}_G)\sqrt{G}|
\right)^{-1}\sqrt{G},
\Label{27-10}
\end{align}
the Holevo bound $C_\theta^H(G)$ can be attained by choosing
a suitable classical data processing.
Thus, (\ref{27-9}) and (\ref{27-10}) can be regarded
as the optimal MSE matrix and the optimal Fisher information matrix
under the weight matrix $G$, respectively.

Especially, 
concerning the D-invariant case,
the equation (\ref{14-3}) guarantees that
the optimal MSE matrix is
\begin{align*}
\re (\tilde{J}_\theta)^{-1} +
\sqrt{G}^{-1} | \sqrt{G} \im  (\tilde{J}_\theta)^{-1}
\sqrt{G}|\sqrt{G}^{-1},
\end{align*} 
and the Fisher information matrix is
\begin{align*}
\sqrt{G}\left(
\sqrt{G} \re (\tilde{J}_\theta)^{-1} \sqrt{G}+
| \sqrt{G} \im (\tilde{J}_\theta)^{-1}\sqrt{G}|
\right)^{-1}\sqrt{G}
\end{align*}
for a given weight matrix $G$.

\subsection{Quantum CR bound}\Label{25-9}
Next, we discuss the asymptotic estimation error
of an estimator based on 
collective measurement on $n$-fold tensor product system
${\cal H}^{\otimes n}\defeq
\overbrace{{\cal H} \otimes \cdots \otimes{\cal H}}^{n}$.
In this case, we treat the estimation problem of
the $n$-fold tensor product family 
${\cal S}^{\otimes n}\defeq \{ \rho_\theta^{\otimes n}\defeq 
\overbrace{\rho_\theta \otimes \cdots \otimes\rho_\theta}^{n}
| \theta \in \Theta\}$.
Then, we discuss the limiting behavior of 
$\tr G V_{\theta}(M^n)$, where
$M^n$ is an estimator of the family of ${\cal S}^{\otimes n}$,
and ${\rm V}_{\theta}(M^n)$ is its MSE matrix.
In the asymptotic setting, we focus on the asymptotically unbiased
conditions (\ref{aub1}) and (\ref{aub2}) 
instead of the locally unbiased condition,
\begin{align}
{\rm E}_{n,\theta}^j &=
{\rm E}_{\theta}^j(M^n)
\defeq\int_{\Omega} \hat{\theta}^j_n 
\Tr M^n(\,d \hat\theta)\rho_\theta^{\otimes n}
\to \theta^j
\Label{aub1}\\
A_{n,\theta;k}^j &= 
A_{\theta;k}^j(M^n)
\defeq 
\frac{\partial}{\partial \theta^k}{\rm E}^j_{\theta}(M^n)
\to \delta^j_k ,
\Label{aub2}
\end{align}
as $n \to \infty$.
Thus, we define the quantum Cram\'{e}r-Rao type bound
(quantum CR bound)
$C_\theta(G)$ as
\begin{align}
& C_\theta(G) \nonumber\\
\defeq &
\min_{\{M^n\}_{n=1}^\infty}
\left\{\left.
\lim_{n \to\infty}
n \tr {\rm V}_{\theta}(M^n)G \right|
\begin{array}{l}
\{ M^n\} \mbox{ is asympto-}\\
\mbox{tically unbiased}
\end{array}
\right\}.\Label{qcrb}
\end{align}

As is independently shown by Hayashi \& Matsumoto\cite{HM} and 
Gill \& Massar\cite{GM},
if the state family satisfies some regularity conditions, {\em e.g.},
continuity, boundedness, {\em etc},
the following two-stage adaptive estimator $M^n$ attains
the bound $\hat{C}_\theta(G)$.
First, we choose a POVM $M$ such that the Fisher information matrix
$J_\theta^M$ is strictly positive for any $\theta \in \Theta$,
and perform it on $\sqrt{n}$ systems.
Then, we obtain the MLE $\hat{\theta}'$ for the family of 
probability distributions $\{ {\rm P}_\theta^M| \theta \in \Theta\}$
based on $\sqrt{n}$ outcomes $\omega_1, \ldots, \omega_{\sqrt{n}}$.
Next, we choose the measurement $M_{\hat{\theta}'}$ which attains
the quasi-quantum Cram\'{e}r-Rao bound $\hat{C}_\theta(G)$,
and perform it on the remaining $n- \sqrt{n}$ systems.
This estimator attains that
$\hat{C}_\theta(G)$, {\em i.e.},
$\tr {\rm V}_\theta(M^n) G \cong \frac{1}{n} \hat{C}_\theta(G)$.
Also, it satisfies the conditions (\ref{aub1}) and (\ref{aub2}).
Therefore, we obtain
\begin{align*}
\hat{C}_\theta(G) \ge C_\theta(G) .
\end{align*}
Moreover, by applying the above statement to
the family ${\cal S}^{\otimes n}$,
we obtain 
\begin{align*}
n \hat{C}_\theta^n(G) \ge C_\theta(G) ,
\end{align*}
where $\hat{C}_\theta^n(G)$ denotes the
quasi-quantum Cram\'{e}r-Rao bound of the family ${\cal S}^{\otimes n}$.

In the $n$-fold tensor product family ${\cal S}^{\otimes n}$,
the SLD $L_{\theta,n;j}$ 
and the RLD $\tilde{L}_{\theta,n;j}$ are given as
\begin{align*}
L_{\theta,n;j}= \sqrt{n} L_{\theta;j}^{(n)} ,\quad
\tilde{L}_{\theta,n;j}= \sqrt{n} \tilde{L}_{\theta;j}^{(n)} ,
\end{align*}
where
\begin{align*}
X^{(n)} &\defeq \frac{1}{\sqrt{n}} \sum_{j=1}^n X^{(n,j)}\\
X^{(n,j)}&\defeq \underbrace{I \otimes \cdots \otimes I}_{j-1}
\otimes X \otimes 
\underbrace{I \otimes \cdots \otimes I}_{n-j}.
\end{align*}
Therefore, 
the SLD Fisher matrix of 
${\cal S}^{\otimes n}$ is calculated as
\begin{align*}
& \Tr \rho^{\otimes n}
(L_{\theta,n;k} \circ L_{\theta,n;j})
= \Tr \rho^{\otimes n}
(\sum_{l=1}^n L_{\theta;k}^{(n,l)}
\circ \sum_{l'=1}^n L_{\theta;j}^{(n,l')}) \\
= &
\sum_{l=1}^n \sum_{l'=1}^n 
\Tr \rho^{\otimes n}
(L_{\theta;k}^{(n,l)} \circ L_{\theta;j}^{(n,l')}) \\
= &
\sum_{l=1}^n 
\Tr \rho^{\otimes n}
(L_{\theta;k}\circ L_{\theta;j})^{(n,l)} \\
& +
\sum_{l=1}^n \sum_{l'\neq l}
\Tr \rho^{\otimes n}
\underbrace{I \cdots I}_{l-1}
\otimes L_{\theta;k} \otimes
\underbrace{I  \cdots  I}_{l'-l}
\otimes L_{\theta;j} \otimes
\underbrace{I\cdots I}_{n-l'} \\
= &
\sum_{l=1}^n 
J_{\theta;k,j} =
n J_{\theta;k,j} .
\end{align*}
Similarly, 
the RLD Fisher matrix of 
${\cal S}^{\otimes n}$ equals
the $n$ times of $\tilde{J}_{\theta}$.
As is shown in Appendix \ref{10-4-7}, 
a similar relation with respect to the Holevo bound
holds as follows.
\begin{lem}\Label{1-1}
Let $C_\theta^{H,n}(G)$ be the
Holevo bound of ${\cal S}^{\otimes n}$,
then
\begin{align}
C_\theta^{H,n}(G) = \frac{1}{n}C_\theta^{H}(G). \Label{8-27-5}
\end{align}
\end{lem}

Thus, 
we can evaluate $C_\theta(G)$
as follows. It proof will be given in Appendix \ref{10-4-8}.
\begin{thm}\Label{th-29}
The quantum CR bound is 
evaluated as
\begin{align}
C_\theta(G) \ge C_\theta^H(G) .\Label{29-1}
\end{align}
\end{thm}

Therefore,
if there exists estimators $M^n$ for $n$-fold tensor product family
${\cal S}^{\otimes n}$ such that 
\begin{align*}
n \tr G V_\theta(M^n) \to C^H_\theta(G),
\end{align*}
then the relation
\begin{align}
C_\theta(G)= C^H_\theta(G)= \lim n\hat{C}_\theta^{n}(G)\Label{27-20}
\end{align}
holds.
Furthermore, 
if the relation (\ref{27-20}) holds in a D-invariant model,
any submodel of it satisfies the relation (\ref{27-20}). 
\subsection{General error function}
In the above discussion, 
we focus only on the trace of the product of the MSE matrix and
a weight matrix.
However, in general, 
we need to take the error function $g(\theta,\hat{\theta})$
other than the above
into consideration.
In this case, similarly to (\ref{qcrb}) we can 
define the asymptotic minimum error $C_\theta(g)$ as
\begin{align*}
& C_\theta(g)\\
\defeq &
\min_{\{M^n\}_{n=1}^\infty}
\left\{
\lim_{n \to\infty}
n {\cal R}^g_\theta(M^n)
\left|
\begin{array}{l}
\{ M^n\} \mbox{ is asympto-}\\
\mbox{tically unbiased}
\end{array}
\right.\right\},
\end{align*}
where
\begin{align*}
{\cal R}^g_\theta(M^n)
\defeq
\int_{\real^d}
g(\theta,\hat{\theta})
\Tr M^n (\,d \hat\theta)\rho_\theta^{\otimes n}.
\end{align*}
We assume that 
when $\hat{\theta}$ is close to $\theta$,
the error function $g$ can be approximated 
by the symmetric matrix $G^g$ 
as follows:
\begin{align*}
g(\hat\theta,\theta)\cong
\sum_{k,l}
G^g_{k,l}(\hat\theta^k- \theta^k)(\hat\theta^l- \theta^l).
\end{align*}

Similarly to subsection \ref{25-9},
if we choose suitable adaptive estimator $M^n$,
the relation
${\cal R}^g_\theta(M^n)\cong 
\frac{1}{n} \hat{C}_\theta(G^g)$ holds.
Thus,
$C_\theta(g)\le  \hat{C}_\theta(G^g)$.
Also, we obtain 
$C_\theta(g)\le  n \hat{C}_\theta^n(G^g)$.

Conversely,
for a fixed $\theta_0$, we choose local chart $\phi(\theta)$ 
at a neighborhood $U_{\theta_0} $ of $\theta_0$
such that
\begin{align*}
g(\theta_0,\theta)= \sum_{k,l}
G^g_{k,l}(\phi^k(\theta)- \phi^k(\theta_0))
(\phi^l(\theta)- \phi^l(\theta_0)),
\end{align*}
for $\forall \theta \in U_{\theta_0} $.
By applying the above discussions to the family
$\{\rho_\theta |\theta \in U_{\theta_0} \}$,
we obtain
\begin{align*}
C_\theta(g)
\ge C_\theta^H(G^g).
\end{align*}

\section{Quantum Gaussian States Family}\Label{Gau}
Next, we review the estimation of expected parameter of
the quantum Gaussian state.
In this case, Yuen \& Lax \cite{YL} derived quasi CR bound for the specific
weight matrix and Holevo\cite{HolP}
did it for arbitrary weight matrix.
This model is essential for the asymptotic analysis of 
quantum two-level system.
In the boson system,
the coherent state with complex amplitude $\alpha$
is described by the coherent vector $
| \alpha ) \defeq e^{-\frac{|\alpha|^2}{2}}
\sum_{n=0}^{\infty} \frac{\alpha^n}{n !}|n \rangle$,
where $n \rangle $ is the $n$-th number vector.
The quantum Gaussian state is given as
\begin{align*}
\rho_{\zeta,\overline{N}}\defeq 
\frac{1}{\pi \overline{N}}
\int_{\complex}
| \alpha )(\alpha|
e^{- \frac{|\alpha-\zeta|^2}{\overline{N}}}\,d \alpha.
\end{align*}
In particular, the relations
\begin{align*}
\rho_{0,\overline{N}}&=
\frac{1}{\overline{N}+1}
\sum_{n=0}^{\infty}
\left(\frac{\overline{N}}{ \overline{N}+1}\right)^n |n \rangle \langle n|,\\
\rho_{\theta,\overline{N}} &=W_{\theta^1,\theta^2}
\rho_{0,\overline{N}}W_{\theta^1,\theta^2}^*
\end{align*}
hold, where $\theta= \frac{1}{\sqrt{2}}(\theta^1+\theta^2i)$ and
$W_{\theta^1,\theta^2}\defeq e^{i  (- \theta^1 P+ \theta^2 Q) }$.
For the estimation of
the family ${\cal S}_{\overline{N}}\defeq \{ \rho_{\theta,\overline{N}}|
\theta = \frac{1}{\sqrt{2}}(\theta^1+ \theta^2 i)\}$,
the following estimator is optimal.
Let $G$ be the weight matrix,
then the matrix $\hat{G}= \sqrt{\det G} G^{-1}$
has the determinant $1$.
We choose the squeezed state $|\phi_{\hat{G}}\rangle \langle \phi_{\hat{G}}|$
such that
\begin{align*}
\left(
\begin{array}{c}
\langle \phi_{\hat{G}}|Q |\phi_{\hat{G}}\rangle \\
\langle \phi_{\hat{G}}|P |\phi_{\hat{G}}\rangle 
\end{array}
\right)
& = 
\left(
\begin{array}{c}
0\\0
\end{array}
\right)\\
\left(
\begin{array}{cc}
\langle \phi_{\hat{G}}|Q^2 |\phi_{\hat{G}}\rangle &
\langle \phi_{\hat{G}}|Q\circ P |\phi_{\hat{G}}\rangle \\
\langle \phi_{\hat{G}}|Q\circ P |\phi_{\hat{G}}\rangle &
\langle \phi_{\hat{G}}| P^2 |\phi_{\hat{G}}\rangle 
\end{array}
\right)
&= \frac{\hat{G}}{2},
\end{align*}
then the relation 
\begin{align}
|\langle \phi_{\hat{G}}, \frac{1}{\sqrt{2}}(\theta^1+ \theta^2i) )|^2
= \exp (- \sum_{k,j} \theta^k ((\hat{G}+I)^{-1})_{k,j}\theta^{j})
\Label{29-8-1}
\end{align}
holds.
The POVM 
\begin{align*}
M_{\hat{G}}(\,d \hat{\theta}^1 \,d \hat{\theta}^2)
\defeq 
W_{\theta^1,\theta^2} 
|\phi_{\hat{G}}\rangle \langle \phi_{\hat{G}}| 
W_{\theta^1,\theta^2} ^*
\frac{\,d \hat{\theta}^1 \,d \hat{\theta}^2}{2\pi}
\end{align*}
satisfies 
the unbiased condition
\begin{align*}
{\rm E}_\theta^i (M_{\hat{G}})= \theta^i.
\end{align*}
Moreover, (\ref{29-8-1}) guarantees that
$\Tr \rho_{\theta,\overline{N}}
M_{\hat{G}}(\,d \hat{\theta}^1 \,d \hat{\theta}^2)$
is the normal distribution with the covariance matrix $
(\overline{N}+ \frac{1}{2})I +\frac{\hat{G}}{2}$.
Therefore, its error can calculated as follows.
\begin{align}
&\tr G {\rm V}_\theta(M_{\hat{G}})=
\tr G ( (\overline{N}+ \frac{1}{2})I +\frac{\hat{G}}{2})\nonumber \\
=& 
(\overline{N}+ \frac{1}{2})\tr G + \frac{1}{2} \tr G 
\sqrt{\det G}G^{-1}\nonumber\\
=&
(\overline{N}+ \frac{1}{2})\tr G + \sqrt{\det G}. \Label{29-8}
\end{align}
For its details, the following theorem holds.
\begin{thm}\Label{16-3-1}
Holevo\cite{HolP}
The POVM $M_{\tilde{G}}$ satisfies 
\begin{align}
&\left( 
\begin{array}{cc}
\int (\hat{\theta}^1)^2 M_{\tilde{G}}(\,d \hat\theta) &
\int \hat{\theta}^1\hat{\theta}^2 M_{\tilde{G}}(\,d \hat\theta) \\
\int \hat{\theta}^1\hat{\theta}^2 M_{\tilde{G}}(\,d \hat\theta) &
\int (\hat{\theta}^2)^2 M_{\tilde{G}}(\,d \hat\theta) \\
\end{array}
\right) \nonumber\\
= &
\left( 
\begin{array}{cc}
Q^2 & Q \circ P  \\
Q \circ P   & P^2
\end{array}
\right)+
\frac{\sqrt{\det G}}{2}G^{-1}\otimes I.
\Label{16-3}
\end{align}
\end{thm}
It is proved in Appendix \ref{19-8}.

Its optimality is showed as follows.
The derivatives can be calculated as
\begin{align*}
\frac{\partial \rho_{\theta,\overline{N}}}{\partial \theta^1}
& = -i [P, \rho_{\theta,\overline{N}}]=
\frac{1}{\overline{N}+ \frac{1}{2}}(Q- \theta^1)\circ \rho_{\theta,\overline{N}}\\
\frac{\partial \rho_{\theta,\overline{N}}}{\partial \theta^2}
& = i [Q, \rho_{\theta,\overline{N}}]=
\frac{1}{\overline{N}+ \frac{1}{2}}(P- \theta^2)\circ \rho_{\theta,\overline{N}}.
\end{align*}
Therefore, we can calculate as
\begin{align*}
L_{\theta,1} &= \frac{1}{\overline{N}+ \frac{1}{2}}(Q- \theta^1) ,\quad
L_{\theta,2} = \frac{1}{\overline{N}+ \frac{1}{2}}(P- \theta^2) \\
J_\theta &=
\left(
\begin{array}{cc}
(\overline{N}+ \frac{1}{2})^{-1} & 0 \\
0 & (\overline{N}+ \frac{1}{2})^{-1} 
\end{array}
\right) ,\\
J_\theta^{-1}
D_\theta J_\theta^{-1} &= 
\left(
\begin{array}{cc}
0 &  1\\
-1 & 0 
\end{array}
\right) ,
\end{align*}
where we use the relation (\ref{29-7}).
Thus, 
since 
\begin{align*}
\tr \left| \frac{1}{2}\sqrt{G}\left(
\begin{array}{cc}
0 & -i \\ 
i & 0 
\end{array}
\right)\sqrt{G}\right|
= 
\sqrt{\det{G}},
\end{align*}
the RLD Fisher information matrix is
\begin{align*}
\tilde{J}_\theta^{-1}=
\left(
\begin{array}{cc}
\overline{N}+ \frac{1}{2} & i/2  \\
-i/2 & \overline{N}+ \frac{1}{2} 
\end{array}
\right). 
\end{align*}
Thus, the RLD bound is calculated as
\begin{align*}
C_\theta^R(G)= (\overline{N}+ \frac{1}{2})\tr G + 
\sqrt{\det G},
\end{align*}
which equals the right hand side of (\ref{29-8}).
Thus, from (\ref{29-9}), we obtain the optimality of $M_{\hat{G}}$,
{\it i.e.}, Yuen, Lax, and Holevo's result:
\begin{align*}
\hat{C}_\theta(G)= (\overline{N}+ \frac{1}{2})\tr G + 
\sqrt{\det G}.
\end{align*}
Furthermore, for the $n$-fold tensor product model 
${\cal S}_{\overline{N}}^{\otimes n}$,
we can define a suitable estimator as follows.
First, we perform the measurement $M_{\hat{G}}$ on the individual system,
and obtain $n$ data $(\theta^1_1, \theta^2_1), \ldots,
(\theta^1_n, \theta^2_n)$.
We decide the estimate as $\hat{\theta}^k \defeq \frac{1}{n}
\sum_{j=1}^n  \theta^k_j$.
In this case, the MSE matrix equals 
$\frac{1}{n}((\overline{N}+ \frac{1}{2})I +\frac{\hat{G}}{2})$.
Therefore, Theorem \ref{th-29} guarantees 
\begin{align*}
C_\theta(G)=\hat{C}_\theta(G) = (\overline{N}+ \frac{1}{2})\tr G + 
\sqrt{\det G},
\end{align*}
which implies that 
there is no advantage for using the quantum correlation 
in the measurement apparatus in the estimation of 
the expected parameter of quantum Gaussian family.

\section{Asymptotic behavior of Spin $j$ system}\Label{27-11}
In this section, we discuss
how the spin $j$ system asymptotically approaches to
the quantum Gaussian state as $j$ goes to infinity.
Accardi and Bach\cite{A-B1,A-B2} focused on 
the limiting behaviour of the $n$-tensor 
product space of spin 1/2,
but we focus on 
that of spin $j$ system.
Let $J_{j,1},J_{j,2},J_{j,3}$ be the standard generators of 
the spin $j$ representation of Lie algebra $\su(2)$.
That is, the representation space ${\cal H}_j$
is spanned by $
|j,m\rangle,m=j,j-1,\ldots,-j+1,-j$,
satisfying
\begin{align*}
J_{j,3}|j,m\rangle= m |j,m\rangle.
\end{align*}
The matrixes $J_{j,\pm}\defeq J_{j,1} \pm i J_{j,2}$ are represented as
\begin{align*}
J_{j,+} |j,m\rangle &= 
\sqrt{(j-m)(j+m+1)}|j,m+1\rangle \\
J_{j,-} |j,m\rangle &= 
\sqrt{(j-m+1)(j+m)}|j,m-1\rangle .
\end{align*}
For any complex $z= x+i y, |z|\,<1$,
we define 
the special unitary matrix 
\begin{align*}
U_z\defeq \left(
\begin{array}{cc}
\sqrt{1- |z|^2} & -z^* \\
z & \sqrt{1- |z|^2} 
\end{array}
\right),
\end{align*}
and denote its representation on ${\cal H}_j$ by
$U_{j,z}$.
The spin coherent vector
$|j,z) \defeq U_{j,z} |j,j\rangle $ 
satisfies 
\begin{align*}
\langle j,m |j,z)  = 
\sqrt{\left(\begin{array}{c}
2j \\ j+m
\end{array}\right)}
\alpha^{(j-m)}(1- |\alpha|^2)^{\frac{j+m}{2}}.
\end{align*}
We also define the state $\rho_{j,p}$
as
\begin{align*}
\rho_{j,p} \defeq 
\frac{1-p}{1-p^{2j+1}}
\sum_{m=-j}^{j} p^{j -m}
|j,m\rangle \langle j,m|.
\end{align*}
Defining the isometry $W_j$
from ${\cal H}_j$ to 
$L^2(\real)$ as
\begin{align*}
W_j:|j,m\rangle \to | j-m\rangle,
\end{align*}
we can regard the space  ${\cal H}_{j}$ as 
a subspace of $L^2(\real)$.
\begin{thm}\Label{th-31}
Under the above imbeding, we obtain the following limiting characterization
\begin{align}
\rho_{j,p} &\to \rho_{0,\frac{p}{1-p}} \Label{31-1}\\
|j,\frac{z}{\sqrt{2j}})(j,\frac{z}{\sqrt{2j}}| &\to |z)(z| \Label{31-2}
\end{align}
in the trace norm topology.
Moreover, when $j$ goes to infinity, the limiting relations
\begin{align}
&\Tr \rho_{j,p}(a-\frac{1}{\sqrt{2j}}J_{j,+})^*
(a-\frac{1}{\sqrt{2j}}J_{j,+})\to 0 \Label{31-3}\\
&\Tr \rho_{j,p}(a^*-\frac{1}{\sqrt{2j}}J_{j,-})^*
(a^*-\frac{1}{\sqrt{2j}}J_{j,-})\to 0 \Label{31-4}\\
&\Tr \rho_{j,p}(Q- \frac{1}{\sqrt{j}}J_x)^2 \to 0 \Label{31-5}\\
&\Tr \rho_{j,p}(P- \frac{1}{\sqrt{j}}J_y)^2 \to 0 \Label{31-6} \\
&\Tr \rho_{j,p} Q^2 \to \Tr \rho_{0,\frac{p}{1-p}} Q^2  \Label{31-11} \\
&\Tr \rho_{j,p} P^2 \to \Tr \rho_{0,\frac{p}{1-p}} P^2  \Label{31-12} \\
&\Tr \rho_{j,p} (Q\circ P) \to \Tr \rho_{0,\frac{p}{1-p}} (Q\circ P) 
\Label{31-13}\\
&\Tr \rho_{j,p}((Q- \frac{1}{\sqrt{j}}J_x)\circ Q)\to 0 \Label{31-7} \\
&\Tr \rho_{j,p}((Q- \frac{1}{\sqrt{j}}J_x)\circ P)\to 0 \Label{31-8} \\
&\Tr \rho_{j,p}((P- \frac{1}{\sqrt{j}}J_y)\circ Q)\to 0 \Label{31-9} \\
&\Tr \rho_{j,p}((P- \frac{1}{\sqrt{j}}J_y)\circ P)\to 0 \Label{31-10} 
\end{align}
hold,
where we abbreviate the isometry $W_j$.
\end{thm}

\section{Estimation in quantum two-level system}\Label{27-15}
Next, we consider 
the estimation problem of $n$-fold tensor product family of
the full parameter model 
${\cal S}_{\rm full}\defeq \{
\rho_{\theta} \defeq \frac{1}{2} I + \sum_{i=1}^3 \theta^i \sigma_i
| \|\theta \|\le 1 \}$ on the Hilbert space $\complex^2$,
where
\begin{align*}
\sigma_1 = \frac{1}{2}
\left(
\begin{array}{cc}
0 & 1\\
1 & 0 
\end{array}
\right),
\sigma_2 = \frac{1}{2}
\left(
\begin{array}{cc}
0 & -i\\
i & 0 
\end{array}
\right),
\sigma_3 = \frac{1}{2}
\left(
\begin{array}{cc}
1 & 0\\
0 & -1 
\end{array}
\right).
\end{align*}
In this parameterization, 
the SLDs at the point $(0,0,r)$ can be expressed as
\begin{align*}
L_{(0,0,r);1}&= 2 \sigma_1 , \quad
L_{(0,0,r);2}= 2 \sigma_2 \\
L_{(0,0,r);3}&= 
\left(
\begin{array}{cc}
\frac{1}{1+r} & 0\\
0 & \frac{-1}{1-r}
\end{array}
\right)
= \frac{1}{1-r^2}( 2 \sigma_3 - r I)
.
\end{align*}
Then, the SLD Fisher matrix $J_{(0,0,r)}$ 
and
RLD Fisher matrix $\tilde{J}_{(0,0,r)}$ 
at the point $(0,0,r)$
can be calculated as
\begin{align}
J_{\theta}&=
\left(
\begin{array}{ccc}
1 & 0 & 0 \\
0 & 1 & 0 \\
0 & 0 & \frac{1}{1- r^2}
\end{array}\right), \quad
\tilde{J}_{\theta}^{-1}=
\left(
\begin{array}{ccc}
1 & -i r & 0 \\
i r & 1 & 0 \\
0 & 0 & 1- r^2
\end{array}\right).\Label{13-2}
\end{align}
We can also check that this model is D-invariant.
The state $\rho_{\theta}$ is described in the notations 
in Section \ref{27-11}, as
\begin{align*}
\rho_{\theta}=
U_{e^{i \psi}\sin \phi/2}
\rho_{1/2,p(\|\theta\|)}
U_{e^{i \psi}\sin \phi/2}^*
\end{align*}
where
$p(r)= \frac{1+r}{1-r}$ and
$\frac{\theta^1+ i \theta^2}{\|\theta\|}=
e^{i \psi}\sin \phi$.

On the other hand, as was proved by 
Nagaoka\cite{Na}, Hayashi\cite{Ha}, Gill \& Massar\cite{GM},
in any model of the quantum two level system, 
the quasi CR bound can be calculated as
\begin{align}
\hat{C}_\theta(G)= \left(\tr 
\sqrt{J_\theta^{-\frac{1}{2}}
G
J_\theta^{-\frac{1}{2}}
}\right)^2 
= \left(\tr 
\sqrt{\sqrt{G}
J_\theta^{-1}
\sqrt{G}}\right)^2 \Label{10-3-1},
\end{align}
where
the second equation follows from the unitary equivalence
between $A A^*$ and $A^* A$.

\subsection{Covariant approach}\Label{Cov}
As the first step of this problem, 
we focus on the risk function $g$ covariant for 
$\SU(2)$.
Then,
the risk function $R(\hat\theta,\theta)$
can be expressed by
$g(\|\hat\theta\|,\|\theta\|, \phi)$,
where $\phi$ is the angle between $\hat{\theta}$ and $\theta$,
{\em i.e.}, $|\langle\hat{\theta}, \theta\rangle|=
\|\hat{\theta}\|\| \theta\|\cos \phi$.
It can be divided into two parts:
\begin{align*}
g(\|\hat\theta\|,\|\theta\|, \phi)=
f_1(\|\hat\theta\|,\|\theta\|)+ f_{2,\|\hat\theta\|,\|\theta\|}(\phi),
\end{align*}
where
\begin{align*}
f_1(\|\hat\theta\|,\|\theta\|) &\defeq g(\|\hat\theta\|,\|\theta\|, 0) \\
f_{2,\|\hat\theta\|,\|\theta\|}(\phi)&\defeq 
g(\|\hat\theta\|,\|\theta\|, \phi)- g(\|\hat\theta\|,\|\theta\|, 0)
\end{align*}
For example,
the square of the Bures' distance is described as
\begin{align*}
&b^2(\rho_\theta,\rho_{\hat\theta})= 
1-F(\rho_\theta,\rho_{\hat\theta})\\
= &\frac{1}{2}
(1- \sqrt{1- \|\theta\|^2}\sqrt{1- \|\hat\theta\|^2}
- \hat{\theta}\cdot \theta)\\
=&
\frac{1}{2}
(1- \sqrt{1- \|\theta\|^2}\sqrt{1- \|\hat\theta\|^2}- 
 \|\hat{\theta}\|\| \theta\|)\\
&+  \frac{1}{2} \|\hat{\theta}\|\| \theta\| (1- \cos \phi).
\end{align*}
This risk function can be approximated as
\begin{align*}
b^2(\rho_\theta,\rho_{\hat\theta})\cong
\frac{1}{4} \sum_{k,l} J_{\theta,k,l} (\theta^k-\hat\theta^k)
(\theta^l-\hat\theta^l).
\end{align*}
Thus, the relations (\ref{29-9}), (\ref{27-12}), and (\ref{13-2}) 
yield that
\begin{align*}
C_{(0,0,r)} (b^2)\ge C_{(0,0,r)}^H (b^2)
= \frac{3+ 2 r}{4}.
\end{align*}
Therefore, the covariance guarantees that
\begin{align*}
C_{\theta} (b^2)\ge \frac{3+ 2 \|\theta\|}{4}
.
\end{align*}

As another example, we can simplify 
the square of the Euclidean distance $\| \theta - \hat\theta\|$ as follows.
\begin{align*}
& \| \theta - \hat\theta\|^2=
\|\hat{\theta}\|^2 + \| \theta\|^2 -
2\|\hat{\theta}\|\| \theta\|\cos \phi\\
=& (\|\hat{\theta}\|- \| \theta\|)^2 
+ 2\|\hat{\theta}\|\| \theta\|(1-\cos \phi).
\end{align*}
Concerning this risk function,
we obtain
\begin{align}
C_{\theta} (I)
\le C_{\theta}^H (I)= 3 + 2 \|\theta\| - \|\theta\|^2. \Label{13-3}
\end{align}

In the following, we construct a suitable estimator 
for the family
${\cal S}_{\rm full}^{\otimes n}$.
When we focus on
the tensor representation on $(\complex^2)^{\otimes n}$
of $\SU(2)$,
we obtain its irreducible decomposition as 
\begin{align*}
(\complex^2)^{\otimes n}&=
\bigoplus_{j=0 \hbox{ or } 1/2}^{n/2}
{\cal H}_j \otimes {\cal H}_{n,j} \\
{\cal H}_{n,j} &\defeq 
\complex^{\genfrac{(}{)}{0pt}{}{n}{n/2-j}
-\genfrac{(}{)}{0pt}{}{n}{n/2-j-1}
}.
\end{align*}
Using this decomposition, 
we perform the projection measurement $E^n= \{E^n_j\}$ on the system
$(\complex^2)^{\otimes n}$,
where $E^n_j$ is the projection to 
${\cal H}_j \otimes {\cal H}_{n,j}$.
Then, we obtain the data $j$ 
and the final state 
$ U_{j,e^{i \psi}\sin \frac{\phi}{2}}
\rho_{j,p(\|\theta\|)}
U_{j,e^{i \psi}\sin \frac{\phi}{2}}^*
\otimes \rho_{{\rm mix},{\cal H}_{n,j}}$
with the probability
 \begin{align*}
&P_{n,\|\theta\|}(j)\\
\defeq &
\left(\genfrac{(}{)}{0pt}{}{n}{\frac{n}{2}-j}
-\genfrac{(}{)}{0pt}{}{n}{\frac{n}{2}-j-1}\right)\\
& \cdot \Bigl(
(\frac{1- \|\theta\|}{2})^{\frac{n}{2}-j}
(\frac{1+ \|\theta\|}{2})^{\frac{n}{2}+j} \\
&\quad + \cdots +
(\frac{1- \|\theta\|}{2})^{\frac{n}{2}+j}
(\frac{1+ \|\theta\|}{2})^{\frac{n}{2}-j} \Bigr)
\\
=&
\left(\genfrac{(}{)}{0pt}{}{n}{\frac{n}{2}-j}
-\genfrac{(}{)}{0pt}{}{n}{\frac{n}{2}-j-1}\right)\\
& \cdot 
(\frac{1- \|\theta\|}{2})^{\frac{n}{2}-j}
(\frac{1+ \|\theta\|}{2})^{\frac{n}{2}+j+1} 
(1- (\frac{1- \|\theta\|}{1+ \|\theta\|})^{\frac{n}{2}+j+1} ),
\end{align*}
where 
$\rho_{{\rm mix},{\cal H}_{n,j}}$ is the completely mixed state
on the space ${\cal H}_{n,j}$.
Next, we take the partial trace with respect to the space $
{\cal H}_{n,j}$,
and perform the covariant measurement:
\begin{align*}
M^j(\phi ,\psi )
\defeq 
(2j+1)
|j ,e^{i \psi}\sin \frac{\phi}{2})(j,e^{i \psi}\sin \frac{\phi}{2}| 
\frac{\sin \phi}{4 \pi} 
\end{align*}
Note that the measure 
$\frac{\sin \phi}{4 \pi} \,d \phi \,d \psi$ 
is the invariant probability measure with parameter 
$\phi\in [0,\pi), \psi\in[0,2\pi)$.
When true parameter is
$(0,0,r)$, 
the distribution of data
can be calculated as
\begin{align*}
& \Tr \rho_{j,p} 
M^j(\phi ,\psi) \\
=&
(2j+1)\frac{1-p}{1-p^{2j+1}}
\left(
1- (1-p)\sin^2 \frac{\phi}{2}
\right)^{2j}
\frac{\sin \phi}{4 \pi}, 
\end{align*}
where $p= p(r)$.

Finally, based on the data $j$ and $(\phi,\psi)$,
we decide the estimate as
\begin{align*}
\hat\theta^1 = \frac{2j}{n}\cos \psi \sin \phi,~
\hat\theta^2 = \frac{2j}{n}\sin \psi \sin \phi,~
\hat\theta^3 = \frac{2j}{n}\cos \phi.
\end{align*}
Hence, our measurement can be described by the POVM
$M_{{\rm cov}}^n\defeq 
\{M^j(\phi , \psi)\otimes I_{{\cal H}_{n,j}}\}$ with the outcome
$(j,\phi,\psi)$.

Next, we discuss
the average error of
the square of the Euclidean distance $\|\theta-\hat\theta\|^2$
except for the origin $(0,0,0)$.
For the symmetry,
we can assume that the true parameter is
$(0,0,r)$.
In this case, the average error
of $\|\theta-\hat\theta\|^2$
equals 
\begin{align}
\sum_{j=0 \hbox{ or } 1/2}^{n/2}
P_n (j)
\left(
\left(
\frac{2j}{n} - r
\right)^2 
+ 
2  r \frac{2j}{n}
F_{j,\frac{1-r}{1+r}}
\right),\Label{23-1}
\end{align}
where
\begin{align*}
&F_{j,p}  \\
\defeq &
\int_0^{2\pi}\int_0^{\pi}
(1- \cos \phi)
\Tr \rho_{j,\frac{1-r}{1+r}} 
M^j(\phi,\psi)\,d \phi \,d \psi \\
= &(2j+1)\frac{1-p}{1-p^{2j+1}}
\times \\
& \int_0^{\pi}(1- \cos \phi)
\left(
1- (1-p)\sin^2 \frac{\phi}{2}
\right)^{2j}
\frac{\sin \phi}{2 } \,d \phi \\
= &\frac{(2j+1)(1-p)}{2(1-p^{2j+1})}
\int_{-1}^{1}
(1- x)
\left(
\frac{1+p}{2}+
\frac{1-p}{2}x
\right)^{2j}
 \,d x \\
=&
\frac{2\left(
1+ (2j+1)p^{2j+2} - (2j+2)p^{2j+1} 
\right)}{(2j+2)(1-p)(1-p^{2j+1})}.
\end{align*}
Thus, for any fixed $p\,< 1$, we have
\begin{align}
F_{j,p}= \frac{1}{1-p} \frac{1}{j+1} + O(j)p^{2j}.
\Label{13-7}
\end{align}
Using the above relation, 
the first and second terms of (\ref{23-1}) can be calculated as
\begin{align}
&\sum_{j=0 \hbox{ or } 1/2}^{n/2}
P_{n,r} (j)
2  r \frac{2j}{n} F_{j,\frac{1-r}{1+r}}
=
\frac{2}{1+r}(
\frac{4 r}{n}- \frac{2}{n^2})+ o(\frac{1}{n^2})\nonumber\\
& \sum_{j=0 \hbox{ or } 1/2}^{n/2}
P_{n,r} (j)
\left(
\frac{2j}{n} - r
\right)^2  \nonumber \\
= &(1-r^2)\frac{1}{n}-
\frac{2(1-r)}{r}\frac{1}{n^2}
+ O ((1-r^2)^{n/2})
. \Label{13-4}
\end{align}
For a proof of (\ref{13-4}), see Appendix \ref{28-1}.

Therefore, the average error can be approximate as
\begin{align*}
& \int \| \hat\theta- \theta\|^2 \Tr M^n_{\rm cov}(d \hat\theta)
\rho_\theta^{\otimes n}\\
=&
(3+2r-r^2)\frac{1}{n}
-
(\frac{2(1-r)}{r}+\frac{4}{1+r})
\frac{1}{n^2}.
\end{align*}
Combining this and (\ref{13-3}), we obtain
\begin{align}
C_{\theta} (I)
= C_{\theta}^H (I)= 3 + 2 \|\theta\| - \|\theta\|^2.
\Label{10-3-2}
\end{align}
The average of the square of the Bures' distance
also can be calculated 
by the use of the relations (\ref{13-7}) and (\ref{13-4}) as
\begin{align*}
&\frac{1}{2}{\rm E}_{P,j}
\left(
1- \sqrt{1-r^2}\sqrt{1-(\frac{2j}{n})^2 }
- \frac{2j}{n} r
+ 
\frac{2jr}{n}
F_{j,\frac{1-r}{1+r}}\right) \\
\cong &
\frac{1}{4(1-r^2)}(1-r^2)
+ \frac{1}{2} (1+r )\frac{1}{n} \\
= &\left(\frac{3}{4}+ \frac{r}{2}\right)
\frac{1}{n} 
= C_\theta^H(b^2)\frac{1}{n} ,
\end{align*}
where we use 
the following approximation
\begin{align*}
1- \sqrt{1-r^2}\sqrt{1-(\frac{2j}{n})^2 }
- \frac{2j}{n} r
\cong  \frac{1}{2(1-r^2)} \left( \frac{2j}{n}- r\right)^2
\end{align*}
for the case when $\frac{2j}{n}$ is close to $r$.
Thus, 
\begin{align*}
C_\theta(b^2)= \frac{3}{4}+ \frac{r}{2}.
\end{align*}
As a byproduct, we see that 
\begin{align*}
\frac{2j}{n} \to r \hbox{ as } n \to \infty
\end{align*}
in probability $P_{n,r}$.

Next, we proceed to the asymptotic behavior at the origin $(0,0,0)$,
In this case the data $j$ obeys the distribution 
$P_{n,0}$:
\begin{align*}
P_{n,0}(j)\defeq 
\frac{1}{2^n}
\left(\genfrac{(}{)}{0pt}{}{n}{\frac{n}{2}-j}
-\genfrac{(}{)}{0pt}{}{n}{\frac{n}{2}-j-1}\right)
(2j+1) .
\end{align*}
As is proved in Appendix \ref{19-2},
the average error of the square of the Euclidean distance 
can be approximated as
\begin{align}
\sum_j P_{n,0}(j)
\left(\frac{2j}{n}\right)^2
\cong 
\frac{3}{n} - \frac{4 \sqrt{2}}{ \sqrt{\pi }n \sqrt{n}}+ \frac{2}{n^2} 
\Label{19-1}.
\end{align}
Since
\begin{align}
\int \| \hat\theta- (0,0,0)\|^2 \Tr M^n_{\rm cov}(d \hat\theta)
\rho_{(0,0,0)}^{\otimes n}
= \sum_j P_{n,0}(j)
\left(\frac{2j}{n}\right)^2,
\end{align}
we obtain
$C_{(0,0,0)}(I)= C_{(0,0,0)}^H(I)= 3$,
{\em i.e.}, the equation (\ref{10-3-2}) holds at the origin $(0,0,0)$.

On the other hand, by using (\ref{10-3-1}), 
the quasi quantum CR bound can 
be calculated
\begin{align}
\hat{C}_\theta (I)= (2 + \sqrt{1- \|\theta\|^2})^2
= 5- \|\theta\|^2+ 4 \sqrt{1- \|\theta\|^2}.
\end{align}
Since $5- \|\theta\|^2+ 4 \sqrt{1- \|\theta\|^2}-
(3+ 2 \|\theta\| - \|\theta\|^2)=
2(1- \|\theta\|) + 4 \sqrt{1- \|\theta\|^2}$ is strictly greater than $0$
in the mixed states case,
using quantum correlation in the measuring apparatus can
improve the estimation error.

\begin{rem}\rm
The equation (\ref{13-7})
gives the asymptotic behavior of the error 
of $M^j(\phi,\psi)$:
$F_{j,p}\cong \frac{1}{(1-p)j}$.
It can be checked from another viewpoint.
First, we focus on another parameterization:
\begin{align*}
M^j(z) \,d z\defeq (2j+1)|j,z)(j,z| d z.
\end{align*}
The equation (\ref{31-2}) of Theorem \ref{th-31}
guarantees that
the POVM $M^j(\frac{z}{\sqrt{2j}}) $ goes to the POVM $|z)(z|$.
Thus, the equation (\ref{31-1}) guarantees that 
its error goes to $0$ with the rate $\frac{1}{(1-p)j}$.
This fact indicates the importance of 
approximation mentioned by Theorem \ref{th-31}.
Indeed, it plays an important role for 
the general weight matrix case.
\end{rem}
\begin{rem}\rm
One may think that 
the right hand side (R.H.S.) of (\ref{13-4})
is strange because it is better than 
$(1-r^2)\frac{1}{n}$, {\em i.e.,}
the error of the 
efficient estimator the binomial distribution.
That is, 
when data $k$ obeys $n$-binomial distribution with 
parameter $(\frac{1-r}{2},\frac{1+r}{2})$ and 
we choose the estimator of $\theta$ as $k/n$ 
(it is called the efficient estimator),
the error equals $(1-r^2)\frac{1}{n}$,
which is larger than the right hand side of (\ref{13-4}).
However, in mathematical statistics, it is known that
we can improve the efficient estimator except for one point
in the asymptotic second order.
In our estimator, the right hand side of (\ref{13-4})
at $r=0$ 
is given in (\ref{19-1}), and
is larger than $(1-r^2)\frac{1}{n}$.
\end{rem}
\subsection{General weight matrix}\Label{Gen}
Next, we proceed to the general weight matrix.
For the $\SU(2)$ symmetry,
we can focus only on the point $(0,0,r)$ without of
loss of generality.
Concerning the RLD bound, we obtain the following lemma.
\begin{lem}\Label{le27-14}
For the weight matrix
$G= \left(
\begin{array}{cc}
\tilde{G} & g \\
g^T & s
\end{array}
\right)$,
the RLD bound at $(0,0,r)$
can be calculated as
\begin{align}
C_{(0,0,r)}^R(G)
=
\tr G - r^2 s
+ 2r \sqrt{\det \tilde{G}}
,\Label{15-1}
\end{align}
where $\tilde{G}$ is a $2\times 2$ symmetric 
matrix and $g$ is a 2-dimensional vector.
\end{lem}
For a proof, see Appendix \ref{le27-14-pf}.
The main purpose of this subsection is 
the following theorem
\begin{thm}\Label{10-3-5}
Assume the same assumption as Lemma \ref{le27-14},
then
\begin{align}
C_{(0,0,r)}(G)
=C_{(0,0,r)}^R(G)=
\tr G - r^2 s
+ 2r \sqrt{\det \tilde{G}}
.\Label{10-3-3}
\end{align}
\end{thm}
Furthermore, 
as is shown in Appendix \ref{10-3-6},
the inequality 
\begin{align}
C_{\theta}(G)= C_{\theta}^R(G) \,< \hat{C}_{\theta}(G)
\Label{10-3-7}
\end{align}
holds.
Thus, using quantum correlation in measuring apparatus
can improve estimation error in the asymptotic setting.

As the first step of our proof of Theorem \ref{10-3-5},
we characterize the MSE matrix attaining the RLD bound $C_{(0,0,r)}^R(G)$.
The matrix
\begin{align*}
{\rm V}_{\tilde{G},r} \defeq &
\left(
\begin{array}{cc}
I+ r \sqrt{\det \tilde{G}}\cdot \tilde{G}^{-1}
& 0\\
0 & 1-r^2
\end{array}
\right)
\end{align*}
satisfies
${\rm V}_{\tilde{G},r} \ge \tilde{J}_{(0,0,r)}^{-1}$ and
\begin{align*}
\tr G {\rm V}_{\tilde{G},r}
= &
\tr G - r^2 s 
+
r \tr \sqrt{\det \tilde{G}}\cdot
\tilde{G}^{-1}\tilde{G}\\
=&
\tr G - r^2 s 
+
2r \sqrt{\det \tilde{G}}
= C_{(0,0,r)}^R(G).
\end{align*}
Thus, when there exists 
a locally unbiased estimator with 
the covariance matrix ${\rm V}_{\tilde{G}}$,
the RLD bound $C_{(0,0,r)}^R(G)$ can be attained.

In the following, 
we construct an estimator $M^n$ locally unbiased at $(0,0,r_0)$
for the $n$-fold tensor product family
${\cal S}_{\rm full}^{\otimes n}$
such that
$n {\rm V}_{(0,0,r_0)}(M^n) \to V_G$.
In the family ${\cal S}_{\rm full}^{\otimes n}$,
the SLDs can be expressed as
\begin{align*}
\sqrt{n} L_{(0,0,r),k}^{(n)}
=2 \sqrt{n} \sigma_k^{(n)}
=
2 \bigoplus_{j}
J_{j,k} \otimes I_{{\cal H}_{n,j}} 
\end{align*}
for $k=1,2$, and
\begin{align*}
\sqrt{n} L_{(0,0,r),3}^{(n)}
&=\frac{1}{1-r^2} \left( 2\sqrt{n} \sigma_3^{(n)}
- rI\right) \\
&=
\frac{1}{1-r^2}\left( \bigoplus_{j}
2 J_{j,3} \otimes I_{{\cal H}_{n,j}} 
- rI_{(\complex^2)^{\otimes n}}\right).
\end{align*}

First, we perform the projection-valued measurement 
$E^n=\{E^n_j\}$.
Based only on this data $j$,
we decide the estimate of the third parameter $\hat{\theta}^3_r$ as
\begin{align}
\hat{\theta}^3_r(j)\defeq
\frac{1}{J_{n,r}}
\frac{\,d \log P_{n,r}(j)}{\,d r}
+r,\Label{17-1}
\end{align}
where
\begin{align*}
J_{n,r}\defeq 
\sum_j P_{n,r}(j)
\left(
\frac{\,d \log P_{n,r}(j)}{\,d r}
\right)^2 .
\end{align*}
Then, we can easily check that this estimator
$\hat\theta^3_r$ satisfies the following conditions:
\begin{align}
\Tr 
\left.\frac{\partial \rho_\theta^{\otimes n}}
{\partial \theta^k}
\right|_{\theta=(0,0,r)}
(\sum_j \hat{\theta}^3_r(j) E^n_j)
&= 
\left\{
\begin{array}{ll}
1 & k=3 \\
0 & k= 1,2
\end{array}
\right. \Label{17-5} \\
\Tr 
\rho_{(0,0,r)}^{\otimes n}
(\sum_j \hat{\theta}^3_r(j) E^n_j)
&=  0 \Label{17-7}.
\end{align}
The definition guarantees the equation (\ref{17-7}) and 
the equation (\ref{17-5}) for $k=3$.
The rest case can be checked as follows.
The derivative of $\rho_{\theta}$ with respect to the first or second
parameter at the point $(0,0,r)$ can be replaced by
the derivative of $U_{x+iy} \rho_{(0,0,r)} U_{x+iy}^*$
with respect to $x$ or $y$.
Since the probability
$\Tr U_{x+iy}^{\otimes n} \rho_{(0,0,r)}^{\otimes n}
(U_{x+iy}^{\otimes n})^* M_j$ is independent of $x+ iy$,
we have
\begin{align}
\frac{\partial \Tr \rho_{\theta}^{\otimes n}E_j^n }
{\partial \theta^k} = 0 \hbox{ for }k= 1,2,
\Label{17-2}
\end{align}
which implies (\ref{17-5}) in the case of $k=1,2$.

Next, we 
take the partial trace with respect to 
${\cal H}_{n,j}$, and
perform the POVM
$V_j^* M_{\tilde{G}} (\,d x^1 \,d x^2)V_j$ on the 
space ${\cal H}_j$.
After this measurement, we decide 
the estimate of the parameters $ \hat{\theta}^1,\hat{\theta}^2$
as
\begin{align*}
\left(
\begin{array}{c}
\hat{\theta}^1\\
\hat{\theta}^2
\end{array}
\right)
= B_{j,r}^{-1} 
\left(
\begin{array}{c}
x^1\\
x^2
\end{array}
\right),
\end{align*}
where
\begin{align*}
B_{j,r}
\defeq 
\left(
\begin{array}{cc}
\Tr (\rho_{j,p} \circ 2J_{j,1}) V_j^* Q V_j &
\Tr (\rho_{j,p} \circ 2J_{j,2}) V_j^* Q V_j  \\
\Tr (\rho_{j,p} \circ 2J_{j,1}) V_j^* P V_j &
\Tr (\rho_{j,p} \circ 2J_{j,2}) V_j^* P V_j  
\end{array}
\right).
\end{align*}
As is shown in Appendix \ref{27-16}, the relations
\begin{align}
\Tr \frac{\partial \rho_{(0,0,r)}^{\otimes n}}
{\partial \theta^k}
\left(\bigoplus_j 
\left(\int_{\real} 
\hat\theta^l
M_{j,\tilde{G}}(\,d \hat\theta)\right)
\otimes I_{{\cal H}_{n,j}}\right) 
&= \delta_k^l \Label{17-4-1}\\
\Tr \rho_{(0,0,r)}^{\otimes n}
\left(\bigoplus_j 
\left(\int_{\real} 
\hat\theta^l
M_{j,\tilde{G}}(\,d \hat\theta)\right)
\otimes I_{{\cal H}_{n,j}}\right) &=0 \Label{17-4-2}
\end{align}
hold for $l=1,2, k=1,2,3$.
Therefore, 
we see that our estimator $(\hat\theta^1,\hat\theta^2,\hat\theta^3_r)$
is locally unbiased at $(0,0,r)$.

Next, we prove that its covariance matrix ${\rm V}_n$ 
satisfies
\begin{align}
{\rm V}_n \cong 
\left(
\begin{array}{cc}
I + r\sqrt{\det \tilde{G}}\tilde{G}^{-1} & 0 \\
0 & 1-r^2
\end{array}
\right)
\frac{1}{n}
= {\rm V}_{\tilde{G},r}\frac{1}{n}
.\Label{16-6}
\end{align}
Using the equation (\ref{16-8}) in Appendix \ref{27-16}, we have
\begin{align*}
\Tr \rho_{(0,0,r)}^{\otimes n}
\left(\bigoplus_j 
\left(\int_{\real} 
\hat\theta^l (\hat\theta^3- r)
M_{j,\tilde{G}}(\,d \hat\theta)\right)
\otimes I_{{\cal H}_{n,j}}\right) 
= 0 
\end{align*}
for $l=1,2$.
The definition of $\hat\theta^3_r (j)$ guarantees that
\begin{align*}
&\Tr \rho_{(0,0,r)}^{\otimes n}
\left(\bigoplus_j 
\sum_j
(\hat\theta^3 (j) - r)^2
M_{j}
\otimes I_{{\cal H}_{n,j}}\right) \\
= & 
\sum_j P_{n,r}(j) 
\left(\frac{1}{J_{r_0}}
\frac{\,d \log P_{n,r}(j)}{\,d r}\right)^2
= J_{n,r}^{-1} .
\end{align*}
As is shown in Appendix \ref{19-4},
the above value can be approximated by
\begin{align}
J_{n,r}^{-1} \cong (1-r^2)\frac{1}{n} 
+ \frac{1-r^2}{r^2}\frac{1}{n^2} \Label{19-3}.
\end{align}
In order to discuss other components of covariance matrix,
we define
 the $2 \times 2$ matrix ${\rm V}_{j,\tilde{G},r} $:
\begin{align*}
 [{\rm V}_{j,\tilde{G},r}^{k,l}] \defeq 
[\Tr \rho_{j,p}
\int x^k x^l V_j M_{\tilde{G}}(\,d x)V_j^*].
\end{align*}
By use of Theorem \ref{16-3-1}, this matrix can be calculated as
\begin{align*}
\left(
\begin{array}{cc}
\Tr \rho_{j,p} V_j Q^2 V_j^* & \Tr \rho_{j,p} V_j (Q\circ P)  V_j^*\\
\Tr \rho_{j,p} V_j (Q\circ P) V_j^* &\Tr \rho_{j,p} V_j P^2  V_j^*
\end{array}
\right)
+ \frac{\sqrt{\det\tilde{G}}}{2}\tilde{G}^{-1},
\end{align*}
then
the covariance matrix of the estimator $M_{j,\tilde{G}}$ on the state
$\rho_{j,p(r)}$ is
\begin{align*}
B_{j,r}^{-1} 
{\rm V}_{j,\tilde{G},r}(B_{j,r}^{-1} )^T.
\end{align*}
Theorem \ref{th-31} and (\ref{16-3})
guarantee that
\begin{align*}
\frac{1}{\sqrt{j}}B_{j,r} &\to \frac{1}{r}I ,\quad
{\rm V}_{j,\tilde{G},r} \to 
\frac{1}{2r}I + \frac{\sqrt{\det\tilde{G}}}{2}\tilde{G}^{-1}
\end{align*}
as $j \to \infty$. Hence,
\begin{align*}
j B_{j,r}^{-1} 
{\rm V}_{j,\tilde{G},r}(B_{j,r}^{-1} )^T
\to 
\frac{r}{2} I + \frac{r^2\sqrt{\det\tilde{G}}}{2}\tilde{G}^{-1}.
\end{align*}
Thus,
the covariance matrix of our estimator $(\hat\theta^1,\hat\theta^2)$
equals
\begin{align*}
\sum_j P_{n,r}(j) B_{j,r}^{-1} 
V_{j,r}(B_{j,r}^{-1} )^T
\cong (I + r\sqrt{\det\tilde{G}}\cdot\tilde{G}^{-1}
)\frac{1}{n}
\end{align*}
because 
the random variable $\frac{2j}{n}$ converges 
to $r$ in probability $P_{n,r}$.
Thus, we obtain (\ref{16-6}).

Concerning the origin $(0,0,0)$,
we can prove
\begin{align}
\frac{1}{n}J_{(0,0,0)}^{M_{{\rm cov}}^n} \to I,\Label{19-5}
\end{align}
which will be proved in Appendix \ref{19-6}.
Therefore the RLD bound at the origin $(0,0,0)$ can 
be attained.
Then, the proof of Theorem \ref{10-3-5} is completed.

\subsection{Holevo bound in submodel}\Label{Sub}
Since the Holevo bound is attained in the asymptotic sense in 
the full model, Theorem \ref{th-29} guarantees that
the Holevo bound can be attained in the asymptotic sense in 
any submodel ${\cal S}=\{\rho_{\theta(\eta)} | \eta \in \Theta
\subset \real^d\}$.
In the following, we calculate the Holevo bound in 
this case.
Since the Holevo bound equals the SLD Fisher information
in the one-dimensional case,
we treat the two-dimensional case in the following.
First, we suppose that
the true state is $\rho_{(0,0,r)}$.
Without loss of generality,
by choosing a suitable coordinate,
we can assume that
the derivatives can be expressed as
\begin{align*}
D_1 \defeq \left.\frac{\partial \rho_{\theta(\eta)}}{\partial \eta^1}
\right|_{\theta=(0,0,r) }&=
\sigma_1 \\
D_2 \defeq \left.\frac{\partial \rho_{\theta(\eta)}}{\partial \eta^2}
\right|_{\theta=(0,0,r) } &=
\cos \phi \sigma_2
+ \sin \phi \sqrt{1-r^2}\sigma_3,
\end{align*}
where $0 \le \phi \le \frac{\pi}{2}$.
In the above assumption, we have the following theorem.
\begin{thm}\Label{27-18}
Assume that the weight matrix $G$ is parameterized as
\begin{align*}
G= \left(
\begin{array}{cc}
g_1 & g_2 \\
g_2 & g_3
\end{array}
\right).
\end{align*}
When $\frac{g_1}{\sqrt{\det G}}\,< \frac{\cos \phi}{r \sin^2 \phi}$,
the Holevo bound $C^H_{(0,0,r)}(G)$ of the above subfamily can
be calculated as
\begin{align*}
C^H_{(0,0,r)}(G)
=\tr G + 2r \cos \phi \sqrt{\det G} - r^2 \sin^2 \phi g_1,
\end{align*}
and can be attained only by the following covariant matrix
${\rm V}_{G}$
\begin{align}
{\rm V}_{G} 
=I 
+r \cos \phi
\sqrt{\det G} \cdot G^{-1}
-
\left(
\begin{array}{cc}
r^2\sin^2 \phi  & 0 \\
0 & 0 
\end{array}
\right).\Label{23-6}
\end{align}
Otherwise, 
the Holevo bound $C^H_{(0,0,r)}(G)$ and 
the covariant matrix $V_{G}$ can be calculated by
\begin{align}
C^H_{(0,0,r)}(G)&=
\tr G + 
\frac{\det G}{g_1}
\left(\frac{\cos \phi}{\sin \phi}\right)^2 \Label{23-7}\\
{\rm V}_{G}&= I+ \frac{\cos^2 \phi}{\sin^2 \phi}
\left(
\begin{array}{cc}
\frac{g_2^2}{g_1^2} & -\frac{g_2}{g_1} \\
-\frac{g_2}{g_1} & 1 
\end{array}
\right) .\Label{23-9}
\end{align}
\end{thm}
For a proof, see Appendix \ref{27-18-1}.

On the other hand, 
the equation (\ref{10-3-1}) guarantees that 
\begin{align*}
\hat{C}_{(0,0,r)}(G)= (\tr \sqrt{G})^2 =
\tr G + 2 \sqrt{\det G }
\end{align*}
in this parameterization because $J_\theta= I$.
Since we can verify the inequality
$\tr G + 2 \sqrt{\det G }
\,> C_{(0,0,r)}^H(G)$ in the above two cases,
we can check 
effectiveness of quantum correlation in the measuring apparatus
in this case.

The set $\{ {\rm V}_G | \det G=1\}$ represents 
the optimal MSE matrixes.
Its diagonal subset equals
\begin{align}
\left\{\left.
I+\left(
\begin{array}{cc}
r t^{-1} \cos \phi - r^2 \sin^2 \phi & 0 \\
0 & r t \cos \phi
\end{array}
\right)\right|
0 \,< t \le \frac{\cos \phi}{r \sin^2 \phi}
\right\}.
\Label{10-6-1}
\end{align}

\section{Discussion}
We proved that 
the estimation error is evaluated 
by the Holevo bound
in the asymptotic setting
for estimators with quantum correlation in the measuring apparatus
as well as for that without quantum correlation.
We construct an estimator attaining the Holevo bound.
In the covariant case, such an estimator is constructed 
as a covariant estimator.
But, in the other case, it is constructed based on 
the approximation of the spin $j$ system 
with sufficient large $j$
to quantum Gaussian states family.

It is also checked based on the previous results
that 
the Holevo bound cannot be attained by 
the individual measurement in the quantum two-level system.
That is, using quantum correlation in the measuring 
apparatus can improve the estimation error in the asymptotic setting
in the quantum two-level system.

Since the full parameter model of the quantum two-level system
is D-invariant,
its Holevo bound equals the the RLD bound.
Thus, its calculation is not so difficult.
However, a submodel is not necessarily 
D-invariant.
Hence, the calculation of its Holevo bound
is not trivial.
By comparing the previous result,
we point out that
this model is different from pure states model
even in the limiting case $r \to 1$.


\section*{Acknowledgment}
The authors are 
indebted to Professor Hiroshi Nagaoka and Professor Akio Fujiwara
for helpful discussions.
They are grateful to Professor Hiroshi Imai for his support and encouragement.

\appendix

\section{Proofs of Theorems and Lemmas in Section \ref{25-1}}\Label{27-3}
\subsection{Proof of Theorem \ref{thm-1}}\Label{10-4-1}
For any complex valued vector $\vec{b}=[b_j]$ and 
we define a complex valued vector $\vec{a}
=[a^j] =J_\theta^{-1} \vec{b}$ and
matrixes
$X_{\vec{b}}\defeq \sum_j X^j b_j$ and
$L_{\vec{a}}\defeq \sum_j L_{\theta;j} a^j$.
Since the assumption guarantees that
$\langle X_{\vec{b}},L_{\vec{a}}  \rangle
=\langle \vec{b},\vec{a}\rangle$,
Schwarz inequality yields that
\begin{align*}
& \langle \vec{b} | Z_\theta (\vec{X})| \vec{b} \rangle
\langle \vec{b} | J_\theta^{-1} | \vec{b} \rangle
=
\langle \vec{b} | Z_\theta (\vec{X})| \vec{b} \rangle
\langle \vec{a} | J_\theta | \vec{a} \rangle \\
=&
\langle X_{\vec{b}}, X_{\vec{b}} \rangle
\langle L_{\vec{a}}, L_{\vec{a}} \rangle
\ge |\langle \vec{b} , \vec{a} \rangle |^2=
|\langle \vec{b} | J_\theta^{-1} | \vec{b} \rangle|^2.
\end{align*}
Therefore,
we obtain
\begin{align*}
\langle \vec{b} | Z_\theta (\vec{X})| \vec{b} \rangle
\ge \langle \vec{b} | J_\theta^{-1} | \vec{b} \rangle,
\end{align*}
which implies (\ref{sld1}).
Similarly we can prove (\ref{rld1}).
\endproof

\subsection{Proof of Lemma \ref{le-B}}\Label{10-4-2}
For any complex valued vector $\vec{b}=[b_j]$,
we define matrix
$X_{\vec{b},M}\defeq \sum_j X^j_M b_j$.
Since 
\begin{align*}
\int_{\real^d} 
 \langle \hat{\theta}, \vec{b} \rangle 
M(\,d \hat\theta)
=X_{\vec{b},M},
\end{align*}
we obtain
\begin{align*}
& \langle \vec{b}| V_\theta(M) \vec{b}\rangle
-
\langle \vec{b}|Z_\theta (\vec{X}_M)| \vec{b}\rangle \\
=&
\int_{\real^d} 
\langle \hat{\theta}, \vec{b} \rangle^*
\langle \hat{\theta}, \vec{b} \rangle 
M(\,d \hat\theta)
- X_{\vec{b},M}^*X_{\vec{b},M} \\
=&
\int_{\real^d} 
( \langle \hat{\theta}, \vec{b} \rangle - X_{\vec{b},M})^*
M(\,d \hat\theta)
( \langle \hat{\theta}, \vec{b} \rangle - X_{\vec{b},M})
\ge 0,
\end{align*}
which implies (\ref{29-3}).
\endproof

\subsection{Proof of Lemma \ref{le8-26}}\Label{10-4-3}
Since the real symmetric matrix $T\defeq  V- \re W $ satisfies
\begin{align*}
T \ge \im W,
\end{align*}
we obtain
\begin{align*}
\tr T &\ge \min\{\tr T' | T':\hbox{ real symmetric},
T \ge \im W \} \\
&=  \tr |\im W|.
\end{align*}
Therefore,
\begin{align*}
& \tr V \ge
\tr \re W+ \tr T 
\ge \tr \re W+ \tr |\im W|.
\end{align*}
\endproof

\subsection{Proof of Theorem \ref{thm9-25}}\Label{10-4-4}
Since 
\begin{align*}
& \rho_\theta \circ L_{\theta;j}= \rho_\theta \tilde{L}_{\theta;j}
= \rho_\theta \circ  \tilde{L}_{\theta;j}+
\frac{i}{2} [ \tilde{L}_{\theta;j}, \rho_\theta] \\
= &\rho_\theta \circ 
\left(  \tilde{L}_{\theta;j}+\frac{i}{2}{\cal D}_\theta(
\tilde{L}_{\theta;j}) \right),
\end{align*}
we have
\begin{align*}
(I+ i \frac{1}{2}{\cal D}_\theta)
(\tilde{L}_{\theta;j})
=L_{\theta;j},
\end{align*}
which implies $\tilde{L}_{\theta;j}=
(I+ i \frac{1}{2}{\cal D}_\theta)^{-1}L_{\theta;j}$.
Since $
\frac{\partial \rho_\theta}{\partial \theta^{j}}
(\frac{\partial \rho_\theta}{\partial \theta^{j}})^*
=(\tilde{L}_{\theta;j})^* \rho_\theta$,
we have
\begin{align*}
& \tilde{J}_{\theta;k,j}
= \Tr \rho_\theta \tilde{L}_{\theta;k}(\tilde{L}_{\theta;j})^*
= \Tr (\tilde{L}_{\theta;j})^* \rho_\theta \tilde{L}_{\theta;k}\\
=  & \Tr \frac{\partial \rho_\theta}{\partial \theta^{j}}
\tilde{L}_{\theta;k} 
= \Tr (\rho_\theta \circ L_{\theta;j})
\tilde{L}_{\theta;k}
= \langle L_{\theta;j}, \tilde{L}_{\theta;k} \rangle_\theta \\
= & \langle L_{\theta;j}, (I+ i \frac{1}{2}{\cal D}_\theta)^{-1}
 L_{\theta;k} \rangle_\theta.
\end{align*}
Next, we define a linear map ${\cal L}$ from $\complex^d$ to $T_\theta$ as follows,
\begin{align*}
\vec{b} \mapsto \sum_j b^j L_{\theta;j},
\end{align*}
then its inverse ${\cal L}^{-1}$ and 
its adjoint ${\cal L}^*$ are expressed as
\begin{align*}
{\cal L}^{-1}&: X \mapsto 
\sum_{k=1}^d (J_\theta^{-1})^{k,j} \langle L_{\theta;k}, X  \rangle_\theta\\
{\cal L}^* &: X \mapsto  \langle L_{\theta;j}, X  \rangle_\theta.
\end{align*}
Thus, the map $\tilde{J}_\theta$ can be described
by 
\begin{align*}
{\cal L}^* \circ (I+ i \frac{1}{2}{\cal D}_\theta)^{-1}
\circ {\cal L} 
= {\cal L}^* \circ P_{T_\theta} (I+ i \frac{1}{2}{\cal D}_\theta)^{-1}
 P_{T_\theta}\circ {\cal L} ,
\end{align*} 
where $P_{T_\theta}$ is the projection to $T_\theta$.
Since $T_\theta$ is invariant for ${\cal D}_\theta$,
\begin{align*}
(P_{T_\theta}(I+ i \frac{1}{2}{\cal D}_\theta)^{-1}P_{T_\theta})^{-1}
=P_{T_\theta}(I+ i \frac{1}{2}{\cal D}_\theta)P_{T_\theta}.
\end{align*}
Therefore,
the inverse 
of $\tilde{J}_\theta$
equals 
\begin{align*}
& {\cal L}^{-1} \circ (P_{T_\theta}(I+ i \frac{1}{2}{\cal D}_\theta)^{-1}P_{T_\theta})^{-1}
\circ ({\cal L}^{*})^{-1} \\
= &{\cal L}^{-1} \circ  P_{T_\theta}
(I+ i \frac{1}{2}{\cal D}_\theta)P_{T_\theta}
\circ ({\cal L}^{-1})^*,
\end{align*}
which implies
\begin{align*}
(\tilde{J}_{\theta}^{-1})^{k,j}
=
\sum_{l,l'}
(J_\theta^{-1})^{k,l} 
\langle L_{\theta;l}, (I+ i \frac{1}{2}{\cal D}_\theta)
L_{\theta;l'}\rangle_{\theta}
(J_\theta^{-1})^{l',j} .
\end{align*}
\endproof

\subsection{Proof of Lemma \ref{le-3}}\Label{10-4-5}
Let $P$ be the projection 
to $T_\theta$ with respect to the inner product
$\langle ~,~\rangle_\theta$,
and $P^c$ be the the projection 
to its orthogonal space with respect to the inner product
When $X$ satisfies the condition (\ref{8-27-1}),
$\langle P (X^k), L_j\rangle_\theta=
\langle  X^k, P (L_j)\rangle_\theta=
\langle  X^k, L_j\rangle_\theta=\delta^k_j$.
Thus, $P(\vec{X})= [P (X^i)]$ satisfies the condition (\ref{8-27-1}).
Moreover,
\begin{align*}
&\Tr \rho_\theta P(X^k) P^c(X^j) \\
= & \Tr 
\left( 
\rho_\theta \circ P(X^k)
+\frac{1}{2}[ \rho , P(X^k)]\right)
P^c (X^j)\\
= & \Tr 
\left( 
\rho_\theta \circ P(X^k)
+\frac{1}{2}\rho \circ {\cal D}_\theta( P(X^k))\right)
P^c (X^j)\\
= & \Tr 
\left( 
\rho_\theta \circ \Bigl( P(X^k)+ \frac{1}{2}
{\cal D}_\theta( P(X^k))\Bigr)\right)
P^c (X^j)\\
=& \langle P(X^k)+ \frac{1}{2}
{\cal D}_\theta( P(X^k)), P^c (X^j)
\rangle_\theta
= 0.
\end{align*}
Thus, we obtain
\begin{align*}
Z_\theta(\vec{X}) = Z_\theta(P(\vec{X}))+ Z_\theta(P^c(\vec{X}))\ge 
Z_\theta(P(\vec{X})),
\end{align*}
which implies that
\begin{align*}
& \sqrt{G} \re Z_\theta(\vec{X}) \sqrt{G} +
| \sqrt{G} \im Z_\theta(\vec{X}) \sqrt{G} |\\
\ge &
\sqrt{G}  Z_\theta(\vec{X}) \sqrt{G} 
\ge \sqrt{G} Z_\theta( P(\vec{X})) \sqrt{G} .
\end{align*}
Since the matrix $\sqrt{G} \im Z_\theta(\vec{X}) \sqrt{G}$
is imaginary Hermite matrix,
$| \sqrt{G} \im Z_\theta(\vec{X}) \sqrt{G} |$
is real symmetric matrix.
Therefore, Lemma \ref{le8-26} guarantees that
\begin{align*}
Z_\theta(\vec{X}) \ge Z_\theta(P(\vec{X}) ),
\end{align*}
which implies (\ref{8-27-2}).

Next, we proceed to a proof of (\ref{10-19-1}).
Since the basis $\langle L_1, \ldots, L_m\rangle$
is normally orthogonal concerning SLD,
the equation (\ref{29-6}) guarantees that
\begin{align}
\Tr \rho_\theta L_k L_j
= \Tr \rho_\theta  L_k \circ L_j
+ \frac{1}{2} \Tr \rho_\theta  [L_k , L_j]
= \delta_{k,j} -i \frac{1}{2}D_{\theta,k,j}
= \tilde{J}_\theta^{-1}.
\end{align}
Hence, when we choose the vector 
$v^k= (v^k_1, \ldots v^k_m)$ satisfying that
$X^k= \sum_{j} v^k_j L_j$,
\begin{align}
\Tr \frac{\partial \rho_\theta}{\partial \theta^k} X^k &= 
\re \langle d_k|J|v^j\rangle  \\
\Tr \rho X^k X^j&= \langle v^k|J|v^j\rangle.
\end{align}
Therefore, we obtain (\ref{10-19-1}).
\endproof

\subsection{Proof of Theorem \ref{thm29-2}}\Label{10-4-6}
There exists
$d_1 \times d_2$ matrix $O$ such that
$\sqrt{P_{\vec{X}}^T G P_{\vec{X}}}=
O \sqrt{G}P_{\vec{X}}$ and
$O^T O= I_{d_1}$.
Since $X^k=
\sum_{l=1}^{d_2} P_{\vec{X};l}^k L_{\theta}^l$,
\begin{align*}
&C_{\theta,1}(G,\vec{X}) \\
=&
\tr \sqrt{G} \re Z_\theta(\vec{X})\sqrt{G}
+ \tr |\sqrt{G} \im Z_\theta(\vec{X})\sqrt{G}| \\
=&
\tr \sqrt{G} P_{\vec{X}} \re Z_\theta(\vec{L})^T 
P_{\vec{X}}\sqrt{G} \\
& + \tr | \sqrt{G} P_{\vec{X}} \im Z_\theta(\vec{L})^T 
P_{\vec{X}}\sqrt{G}| \\
=&
\tr O \sqrt{G} P_{\vec{X}} \re Z_\theta(\vec{L})^T 
P_{\vec{X}}\sqrt{G}^T O \\
&+ \tr | O \sqrt{G} P_{\vec{X}} \im Z_\theta(\vec{L})^T 
P_{\vec{X}}\sqrt{G}^T O| \\
=&
C_{\theta,2}( P_{\vec{X}}^T G P_{\vec{X}},\vec{L}) =
C_{\theta,2}^R(P_{\vec{X}}^T G P_{\vec{X}}).
\end{align*}

Let $\{M^n\}$ be a sequence of locally unbiased estimators
of ${\cal S}_2$
such that
$\Tr V_\theta(M^n) P_{\vec{X}}^T G P_{\vec{X}}
\to C_{\theta,2}^R(^T P_{\vec{X}} G P_{\vec{X}})$.
Next, we define
an estimator ${M^n}'\defeq (P_{\vec{X}},M^n)$ on ${\cal S}_1$
satisfying locally unbiasedness condition at $\theta$.
Its covariance matrix is 
$V_\theta({M^n}')=
P_{\vec{X}} V_\theta(M^n)P_{\vec{X}}^T$.
Hence,
\begin{align*}
&\tr  V_\theta({M^n}')G = \tr V_\theta(M^n) P_{\vec{X}}^T G P_{\vec{X}}\\
\to  &C_{\theta,2}^R(^T P_{\vec{X}} G P_{\vec{X}})
=
C_{\theta,1}(G,\vec{X}).
\end{align*}
\endproof

\subsection{Proof of Lemma \ref{1-1}}\Label{10-4-7}
Let $\overline{T}_\theta^n$ be
the linear space spanned by the orbit of 
the SLD tangent space of ${\cal S}^{\otimes n}$.
Since 
any element $X$ of $T_\theta$ satisfies
\begin{align*}
\sqrt{n}\Bigl(\overbrace{{\cal D}_\theta\circ \cdots \circ 
{\cal D}_\theta}^{k}(X)\Bigr)^{(n)}= 
\sqrt{n} \overbrace{{\cal D}_\theta\circ \cdots \circ 
{\cal D}_\theta}^{k}(X^{(n)}),
\end{align*}
the $\overline{T}_\theta^n$ equals
\begin{align*}
\{ \sqrt{n} X^{(n)}| X \in \overline{T}_\theta\}.
\end{align*}
Furthermore,
the vector $\sqrt{n} \vec{X}^{(n)}= [\sqrt{n} (X^i)^{(n)}]$
satisfies
\begin{align*}
C_\theta (G,\sqrt{n} \vec{X}^{(n)}) 
= n C_\theta (G,\vec{X}).
\end{align*}
Therefore, Lemma \ref{le-3} guarantees that
\begin{align*}
&C_\theta^{H,n}(G)\\
= &\min_{\vec{X}:X^j \in \overline{T}_\theta}
\left\{C_\theta (G,\sqrt{n} \vec{X}^{(n)}) \left|
\langle \sqrt{n} L_{\theta;k}^{(n)},\sqrt{n}(X^j)^{(n)}\rangle_\theta
= \delta_k^j
\right.\right\}\\
= &\min_{X:X^j \in \overline{T}_\theta}
\left\{n C_\theta (G,\vec{X}) \left|
n \langle L_{\theta;k}, X^j \rangle_\theta
= \delta_k^j
\right.\right\}\\
= &\min_{\vec{Y}:Y^j \in \overline{T}_\theta}
\left\{\frac{1}{n} C_\theta (G,\vec{Y}) \left|
\langle L_{\theta;k}, Y^j \rangle_\theta
= \delta_k^j
\right.\right\},
\end{align*}
where we put $\vec{Y}= \frac{1}{n}\vec{X}$.
Therefore, we obtain (\ref{8-27-5}).
\endproof

\subsection{Proof of Theorem \ref{th-29}}\Label{10-4-8}
Lemma \ref{le-B} guarantees that
\begin{align*}
{\rm V}_\theta(M^n) \ge Z_\theta (\vec{X}_{M^n}).
\end{align*}
Since the vector $\vec{Y}_M= (Y_M^i\defeq \sum_j 
(A_\theta(M)^{-1})^i_j X_{M}^j$
of Hermitian matrixes
satisfies 
\begin{align*}
\langle \sqrt{n}L_{\theta;i}, Y_{M^n}^j\rangle_\theta &= \delta_i^j, \\
Z_\theta (\vec{X}_{M^n}) &=
A_\theta(M^n) Z_\theta (\vec{Y}_{M^n}) 
 A_\theta^T(M^n),
\end{align*}
the relations
\begin{align*}
& \tr G {\rm V}_\theta(M^n)\\
\ge &
\tr \sqrt{G} \re A_\theta(M^n) Z_\theta (\vec{Y}_{M^n}) 
A_\theta^T(M^n) \sqrt{G} \\
& + 
\tr| \sqrt{G} \im A_\theta(M^n) Z_\theta (\vec{Y}_{M^n}) 
A_\theta^T(M^n) \sqrt{G}| \\
\ge &
C_\theta^{H,n}(
A_\theta^T(M^n) G A_\theta(M^n))\\
= &
\frac{1}{n}C_\theta^H(
A_\theta^T(M^n) G A_\theta(M^n))
\end{align*}
hold.
Taking the limit, we obtain
\begin{align*}
\lim_{n \to \infty} n \tr G {\rm V}_\theta(M^n)\ge &
\lim_{n \to \infty} C_\theta^H(
 A_\theta^T(M^n) G A_\theta(M^n)) \\
=& C_\theta^H(G ),
\end{align*}
which implies (\ref{29-1}).

\endproof

\section{Proof of Theorem \ref{16-3-1}}\Label{19-8}
Let $E$ be the joint measurement of $P\otimes I + I \otimes P$
and $Q\otimes I - I \otimes Q$ on the space $L^2(\real)\otimes L^2(\real)$.
As was proved in Holevo \cite{HolP},
the POVM $M_{\hat{G}}$ satisfies 
\begin{align}
\Tr M_{\hat{G}}(\,d x \,d y) \rho=
\Tr E (\,d x \,d y) (\rho \otimes 
|\phi_{\hat{G}}\rangle \langle\phi_{\hat{G}}|).
\end{align}
Thus,
\begin{align*}
&\Tr x^2 M_{\tilde{G}}(\,d x \,d y) \rho=
\Tr x ^2 E (\,d x \,d y) 
(\rho \otimes 
|\phi_{\hat{G}}\rangle \langle\phi_{\hat{G}}|)\\
=&
\Tr (Q^2 \otimes I + I \otimes Q^2)
(\rho \otimes 
|\phi_{\hat{G}}\rangle \langle\phi_{\hat{G}}|) 
=
(\Tr Q^2 \rho ) + \hat{G}^{1,1},
\end{align*}
which implies equation (\ref{16-3}) regarding 
the $(1,1)$ element.

Since $\langle\phi_{\hat{G}}|P |\phi_{\hat{G}}\rangle =
\langle\phi_{\hat{G}}|Q |\phi_{\hat{G}}\rangle =0$,
Concerning $(1,2)$ element, 
we have
\begin{align*}
& \Tr x y M_{\tilde{G}}(\,d x \,d y) \rho=
\Tr x y E (\,d x \,d y) 
(\rho \otimes 
|\phi_{\hat{G}}\rangle \langle\phi_{\hat{G}}|) \\
= &
\Tr (Q\circ P \otimes I + I \otimes Q\circ P
- P \otimes Q + Q \otimes P)\\
& \cdot
(\rho \otimes 
|\phi_{\hat{G}}\rangle \langle\phi_{\hat{G}}|) \\
=&
(\Tr (Q\circ P) \rho ) + 
\langle\phi_{\hat{G}}|Q\circ P |\phi_{\hat{G}}\rangle 
=
\Tr (Q\circ P) \rho  + \hat{G}^{1,2}.
\end{align*}
We can similarly prove equation (\ref{16-3}) for other elements.

\section{Proof of Theorem \ref{th-31}}
First, we prove (\ref{31-1}).
Since
\begin{align*}
& \rho_{j,p}- \rho_{0,\frac{p}{1-p}} \\
= &
\frac{p^{2j+1}}{1- p^{2j+1}}
(1-p)
\sum_{n=0}^{2j}
p^n |n \rangle  \langle n | 
 -
(1-p) \sum_{n=2j+1}^{\infty}
p^n |n \rangle  \langle n |,
\end{align*}
we have
\begin{align*}
&\| \rho_{j,p}- \rho_{0,\frac{p}{1-p}}\| \\
=& \frac{p^{2j+1}}{1- p^{2j+1}}
(1-p)
\sum_{n=0}^{2j}p^n 
+ (1-p) \sum_{n=2j+1}^{\infty}
p^n \\
\le &
\frac{p^{2j+1}}{1- p^{2j+1}} + p^{2j+1} \to 0,
\end{align*}
which implies (\ref{31-1}).
Next, we prepare a lemma for our proof of (\ref{31-2}).
\begin{lem}\Label{25-7}
Assume that a sequence of normalized vector 
$a^n= \{a_i^n\}_{i=0}^\infty$ and a normalized vector 
$a= \{a_i\}_{i=0}^\infty$ satisfies
\begin{align*}
a_i^n \to a_i \hbox{ as }n \to \infty,
\end{align*}
then
\begin{align*}
\sum_{i=0}^{\infty} | a_i^n - a_i|^2 \to 0.
\end{align*}
\end{lem}
\begin{proof}
For any real number $\epsilon \,>0 $,
there exists an integers $N_1$ such that
\begin{align*}
\sum_{i=N_1}^\infty |a_i|^2 \le \epsilon.
\end{align*}
Furthermore, we can choose another integer $N_2$ such that
\begin{align*}
\sum_{i=0}^{N_1-1} | a_i^n - a_i | ^2 \,< \epsilon, \quad
\sum_{i=0}^{N_1-1} \left|| a_i^n|^2 - |a_i | ^2 \right|
\,< \epsilon, \quad
\forall n \ge N_2.
\end{align*}
Hence, we have
\begin{align*}
\sum_{i=N_1}^\infty | a_i^n|^2 = 
1- \sum_{i=0}^{N_1-1} |a_i^n|^2
\le 1 - 
\left(
\sum_{i=0}^{N_1-1} |a_i|^2 - \epsilon \right)
\le 2 \epsilon.
\end{align*}
Therefore,
\begin{align*}
& \sum_{i=0}^\infty | a_i^n- a_i|^2 
\le  \sum_{i=0}^{N_1-1} | a_i^n- a_i|^2 
+ 2 \sum_{i=N_1}^\infty (| a_i^n|^2 +| a_i|^2 )\\
\le & \epsilon + 2 (2 \epsilon+ \epsilon )= 7 \epsilon .
\end{align*}
Then, our proof is completed.
\end{proof}

We can calculate $| j, \frac{z}{\sqrt{2j}})$ as
\begin{align*}
&| j, \frac{z}{\sqrt{2j}}) \\
=& \sum_{n=0}^{2j}
\sqrt{\genfrac{(}{)}{0pt}{}{2j}{2j-n}
}
\left(\frac{\alpha}{\sqrt{2j}}\right)^n
\left(1- \frac{|\alpha|^2}{2j}\right)^{\frac{2j-n}{2}}
|n \rangle 
\end{align*}
Its coefficient converges as
\begin{align*}
& \sqrt{\genfrac{(}{)}{0pt}{}{2j}{2j-n}
}
\left(\frac{\alpha}{\sqrt{2j}}\right)^n
\left(1- \frac{|\alpha|^2}{2j}\right)^{\frac{2j-n}{2}}\\
= &
\sqrt{\frac{(2j)!}{(2j-n)!(2j)^n}}
\left( 1- \frac{|\alpha|^2}{2j}\right)^{-n/2}\!\!
\!\!\left(
1- \frac{|\alpha|^2}{2j}
\right)^{\frac{2j}{|\alpha|^2}\cdot \frac{|\alpha|^2}{2}}
\!\!\!\!
\frac{\alpha^n }{\sqrt{n!}} \\
\to & e^{- \frac{|\alpha|^2}{2}} \frac{\alpha^n }{\sqrt{n!}}
\hbox{ as } j \to \infty.
\end{align*}
Thus, Lemma \ref{25-7} guarantees that
\begin{align*}
\left\| |z)- | j, \frac{z}{\sqrt{2j}})\right\| \to 0 ,
\end{align*}
which implies (\ref{31-2}).

\begin{align*}
J_{j,+}| n \rangle &= \sqrt{n}\sqrt{2j-n+1}| n-1 \rangle 
\quad (n= 1, \ldots, 2j)\\
J_{j,+}| 0 \rangle &= 0 \\
J_{j,-}| n \rangle &= \sqrt{n+1}\sqrt{2j-n}| n+1 \rangle 
\quad (n= 0, \ldots, 2j-1)\\
J_{j,-}| 2j \rangle &= 0
\end{align*}

\begin{align*}
& (a-\frac{1}{\sqrt{2j}}J_{j,+})
\rho_{j,p}
(a-\frac{1}{\sqrt{2j}}J_{j,+})^* \\
= &
\frac{1-p}{1- p^{2j+1}}
\sum_{n=1}^{2j}
\Bigl(\sqrt{n}
\Bigl(\sqrt{\frac{2j-n+1}{2j}}-1 \Bigr)\Bigr)^2
p^n
|n-1\rangle \langle n-1|
\end{align*}
Since 
the inequality $1- \sqrt{1-x} \le \sqrt{x}$ holds for
$ 0 \le x \le 1$,
we have
\begin{align*}
& \Tr (a-\frac{1}{\sqrt{2j}}J_{j,+})
\rho_{j,p}
(a-\frac{1}{\sqrt{2j}}J_{j,+})^* \\
\le &
\frac{1-p}{1- p^{2j+1}}
\sum_{n=1}^{2j}
\frac{n (n-1)}{2j}p^n \\
\le &
(1-p)
\sum_{n=1}^{\infty}
\frac{n^2}{2j}p^n
= \frac{1-p}{2j}
\frac{p(1+p)}{(1-p)^3} \to 0,
\end{align*}
which implies (\ref{31-3}).
\begin{align*}
& (a^*-\frac{1}{\sqrt{2j}}J_{j,-})
\rho_{j,p}
(a^*-\frac{1}{\sqrt{2j}}J_{j,-})^* \\
= &
\frac{1-p}{1- p^{2j+1}}
\sum_{n=0}^{2j-1}
\Bigl(\sqrt{n+1}
\Bigl(\sqrt{\frac{2j-n}{2j}}-1 \Bigr)\Bigr)^2
p^n
|n+1\rangle \langle n+1|\\
& +
\frac{1-p}{1- p^{2j+1}}
(2j+1)^2 p^{2j}
|2j+1\rangle \langle 2j+1|.
\end{align*}
Since 
the inequality $1- \sqrt{1-x} \le \sqrt{x}$ holds for
$ 0 \le x \le 1$,
we have
\begin{align*}
& \Tr 
(a^*-\frac{1}{\sqrt{2j}}J_{j,-})
\rho_{j,p}
(a^*-\frac{1}{\sqrt{2j}}J_{j,-})^*  \\
\le &
\frac{1-p}{1- p^{2j+1}}
\sum_{n=0}^{2j-1}
\frac{(n+1)n}{2j}p^n
+ 
\frac{1-p}{1- p^{2j+1}}
(2j+1)^2 p^{2j} \\
\le &
(1-p)
\sum_{n=0}^{\infty}
\frac{(n+1)^2}{2j}p^n
+ 
\frac{1-p}{1- p^{2j+1}}
(2j+1)^2 p^{2j} \\
= &\frac{1-p}{2j}
\frac{1+p}{(1-p)^3} 
+ 
\frac{1-p}{1- p^{2j+1}}
(2j+1)^2 p^{2j}
\to 0,
\end{align*}
which implies (\ref{31-4}).
Since
\begin{align*}
& (Q- \frac{1}{\sqrt{j}}J_x)^2
+(P- \frac{1}{\sqrt{j}}J_y)^2 \\
= &
(a-\frac{1}{\sqrt{2j}}J_{j,+})^*
(a-\frac{1}{\sqrt{2j}}J_{j,+}) \\
&+
(a-\frac{1}{\sqrt{2j}}J_{j,+})
(a-\frac{1}{\sqrt{2j}}J_{j,+})^*,
\end{align*}
the relations (\ref{31-3}) and (\ref{31-4}) guarantee
the relation (\ref{31-5}).
Also, we obtain (\ref{31-6}).

\begin{align*}
&|\Tr \rho_{j,p} Q^2 - \Tr \rho_{0,\frac{p}{1-p}} Q^2 |
+
|\Tr \rho_{j,p} P^2 - \Tr \rho_{0,\frac{p}{1-p}} P^2 |\\
=&
|\Tr (\rho_{j,p} - \rho_{0,\frac{p}{1-p}}) Q^2 |
+
|\Tr (\rho_{j,p} - \rho_{0,\frac{p}{1-p}}) P^2 |\\
=&
\Bigl|\Tr 
\Bigl(\frac{p^{2j+1}}{1- p^{2j+1}}
(1-p)
\sum_{n=0}^{2j}
p^n |n \rangle  \langle n |  \\
& -
(1-p) \sum_{n=2j+1}^{\infty}
p^n |n \rangle  \langle n |\Bigr)
(Q^2+P^2)\Bigr| \\
\le &
\Bigl|\Tr 
\frac{p^{2j+1}}{1- p^{2j+1}}
(1-p)
\sum_{n=0}^{2j}
p^n |n \rangle  \langle n | (Q^2+P^2) \Bigr|\\
& +
\Bigl|\Tr (1-p) \sum_{n=2j+1}^{\infty}
p^n |n \rangle  \langle n |(Q^2+P^2)\Bigr|\\
= &
\frac{p^{2j+1}}{1- p^{2j+1}}
(1-p)\sum_{n=0}^{2j}
p^n (2n+1) \\
&+(1-p) \sum_{n=2j+1}^{\infty}
p^n (2n+1) \\
\le &
p^{2j+1}(1-p)
(\frac{1}{1-p} + \frac{2p}{(1-p)^2})\\
&+ (1-p) \sum_{n=2j+1}^{\infty}
p^n (2n+1) \\
\to &0 \hbox { as } j \to \infty,
\end{align*}
because $\sum_{n=1}^{\infty}p^n (2n+1)\,< \infty$.
Thus, we obtain (\ref{31-11}) and (\ref{31-12}).
\begin{align*}
&|\Tr \rho_{j,p} (Q\circ P) -\Tr \rho_{0,\frac{p}{1-p}} (Q\circ P) |\\
\le &
\Bigl|\Tr 
\Bigl(
\frac{p^{2j+1}}{1- p^{2j+1}}
(1-p)
\sum_{n=0}^{2j}
p^n |n \rangle  \langle n | \\
& -
(1-p) \sum_{n=2j+1}^{\infty}
p^n |n \rangle  \langle n |
\Bigr)
(Q\circ P)\Bigr| \\
\le &
|\Tr 
\frac{p^{2j+1}}{1- p^{2j+1}}
(1-p)
\sum_{n=0}^{2j}
p^n |n \rangle  \langle n | (Q\circ P)| \\
& +|\Tr 
(1-p) \sum_{n=2j+1}^{\infty}
p^n |n \rangle  \langle n |(Q\circ P)|.
\end{align*}
Since
\begin{align*}
-\frac{1}{2}(Q^2 +P^2)
 \le  Q\circ P\le \frac{1}{2}(Q^2 +P^2),
\end{align*}
\begin{align*}
& |\Tr 
(1-p) \sum_{n=2j+1}^{\infty}
p^n |n \rangle  \langle n |(Q\circ P)| \\
\le &
\Tr (1-p) \sum_{n=2j+1}^{\infty}
p^n |n \rangle  \langle n |
\frac{1}{2} (Q^2 +P^2)
\to 0
\end{align*}
Similarly, we ca show
\begin{align*}
\left|\Tr 
\frac{p^{2j+1}}{1- p^{2j+1}}
(1-p)
\sum_{n=0}^{2j}
p^n |n \rangle  \langle n | (Q\circ P)\right|\to 0.
\end{align*}
Thus, we obtain (\ref{31-13}).
By using Schwarz inequality of the inner product
$(X,Y) \mapsto \Tr \rho_{j,p} (X\circ Y)$,
we obtain
\begin{align*}
& \left|\Tr \rho_{j,p}((Q- \frac{1}{\sqrt{j}}J_x)\circ Q)\right|^2 \\
\le &
\Tr \rho_{j,p}(Q- \frac{1}{\sqrt{j}}J_x)^2
\Tr \rho_{j,p} Q^2.
\end{align*}
Thus, the relations (\ref{31-5}) and (\ref{31-11}) guarantee
the relation (\ref{31-7}).
Similarly, we obtain (\ref{31-8}) -- (\ref{31-10}).

\section{Useful formula for Fisher information}
In this section, we explain a useful formula for Fisher information,
which are applied to our proof of (\ref{19-3}) and (\ref{19-5}).
Let ${\cal S}= \{ p_\theta(\omega_1,\omega_2)| \theta \in \Theta \subset \real\}$ be 
a family of probability distributions on $\Omega_1\times \Omega_2$.
We define the marginal distribution $p_\theta(\omega_1)$
and conditional distribution as
\begin{align*}
p_\theta(\omega_1)\defeq \sum_{\omega_2 \in \Omega_2 }
p_\theta(\omega_1,\omega_2) ,\quad
p_\theta(\omega_2|\omega_1)\defeq 
\frac{p_\theta(\omega_1,\omega_2)}{p_\theta(\omega_1)}.
\end{align*}
Then, the following theorem holds for 
the family of distributions
${\cal S}$,
${\cal S}_1 \defeq \{ p_\theta(\omega_1)| 
\theta \in \Theta \subset \real \}$, 
and ${\cal S}_{\omega_1} \defeq \{ p_\theta(\omega_2|\omega_1)| 
\theta \in \Theta \subset \real \}$.
\begin{thm}\Label{29-1-1}
The Fisher information $J_\theta$ of the family ${\cal S}$
satisfies 
\begin{align}
J_\theta= J_{1,\theta}+ \sum_{\omega_1 \in \Omega_1}
p_\theta (\omega_1) J_{\omega_1,\theta}, \Label{19-9}
\end{align}
where
$J_{1,\theta}$ is the Fisher information of ${\cal S}_1$
and $J_{\omega_1,\theta}$ is the Fisher information of ${\cal S}_{\omega_1}$.
Moreover, the information less has another form:
\begin{align}
J_{\omega_1,\theta}
= &
\sum_{\omega_2 \in \Omega_2}
p_\theta(\omega_2|\omega_1)
\left(\frac{\,d \log p_\theta(\omega_2,\omega_1)}{\,d \theta}\right)^2 
\nonumber\\
&- 
\left(
\sum_{\omega_2 \in \Omega_2}
p_\theta(\omega_2|\omega_1)
\frac{\,d \log p_\theta(\omega_2,\omega_1)}{\,d \theta}
\right)^2. \Label{29-2}
\end{align}
\end{thm}
Thus, the average $
\sum_{\omega_1 \in \Omega_1}
p_\theta (\omega_1) J_{\omega_1,\theta}$ can be regarded as
information loss by losing the data $\omega_2$.

\begin{proof}
The Fisher information $J_\theta$ equals
\begin{align*}
&\sum_{\omega_1 \in \Omega_1}\sum_{\omega_2 \in \Omega_2}
p_\theta(\omega_1)p_\theta(\omega_2|\omega_1)
\left(
\frac{\,d \log p_\theta(\omega_1)p_\theta(\omega_2|\omega_1)}{\,d \theta}
\right)^2\\
=&
\sum_{\omega_1 \in \Omega_1}p_\theta(\omega_1)
\sum_{\omega_2 \in \Omega_2}p_\theta(\omega_2|\omega_1) \\
&\times
\left(
\frac{\,d \log p_\theta(\omega_1)}{\,d \theta}
+ \frac{\,d \log p_\theta(\omega_2|\omega_1)}{\,d \theta}
\right)^2\\
=&
\sum_{\omega_1 \in \Omega_1}p_\theta(\omega_1)
\Biggl( \left(\frac{\,d\log p_\theta(\omega_1)}{\,d \theta}
\right)^2 \\
& + \sum_{\omega_2 \in \Omega_2}
p_\theta(\omega_2|\omega_1) \\
&\times 
\left(
\frac{\,d \log p_\theta(\omega_2|\omega_1)}{\,d \theta}
\right)^2
+2 \frac{\,d \log p_\theta(\omega_2|\omega_1)}{\,d \theta}
\frac{\,d\log p_\theta(\omega_1)}{\,d \theta} \Biggr).
\end{align*}
However, the second term is vanished as 
\begin{align*}
&\sum_{\omega_1 \in \Omega_1}\sum_{\omega_2 \in \Omega_2}
p_\theta(\omega_2|\omega_1)p_\theta(\omega_1)
\frac{\,d \log p_\theta(\omega_2|\omega_1)}{\,d \theta}\\
=&\frac{\,d\log p_\theta(\omega_1)}{\,d \theta}
\sum_{\omega_1 \in \Omega_1}p_\theta(\omega_1)
\frac{\,d\log p_\theta(\omega_1)}{\,d \theta}
\sum_{\omega_2 \in \Omega_2}
\frac{\,d p_\theta(\omega_2|\omega_1)}{\,d \theta}\\
=& 0 .
\end{align*}
Thus, we obtain (\ref{19-9}).
Moreover, we can easily check (\ref{29-2}).
\end{proof}
\section{Proofs for Section \ref{27-15}}
\subsection{proof of (\ref{13-4})}\Label{28-1}
The L.H.S. of (\ref{13-4}) can be calculated as
\begin{align*}
& \sum_{j=0 \hbox{ or } 1/2}^{n/2}
P_{n,r} (j)
\left(
\frac{2j}{n} - r
\right)^2  \\
= & 4 \sum_{k=0}^{[n/2]}
P_{n,r} (\frac{n}{2} -k)
\left(
\frac{k}{n} - q(r)\right)^2  \\
=& 4 \left(
q(r)^2 +
\sum_{k=0}^{[n/2]}
P_{n,r} (n/2 -k)
\left(\frac{k^2}{n^2} - 2q(r)\frac{k}{n}\right)
\right),
\end{align*}
where $q(r)\defeq \frac{1-r}{2}$.
Since the probability $P_{n,r} (\frac{n}{2} -k)$ has
another expression:
$P_{n,r} (\frac{n}{2} -k)
=\frac{1}{r}
\left(\genfrac{(}{)}{0pt}{}{n}{k}-
\genfrac{(}{)}{0pt}{}{n}{k-1}\right)
q(r)^k (1-q(r))^{n-k+1}
\left(1- \left(\frac{1-r}{1+r}\right)^{n-k}\right)
$, we can calculate the expectations of $k$ and $k^2$
as follows.
\begin{align*}
&\sum_{k=0}^{[n/2]}
k^2 P_{n,r} (\frac{n}{2} -k) \\
=&
\sum_{k=0}^{[n/2]}
k^2 \frac{1}{r}
\left(\genfrac{(}{)}{0pt}{}{n}{k}-
\genfrac{(}{)}{0pt}{}{n}{k-1}\right)
q(r)^k (1-q(r))^{n-k+1} \\
&+ O\left(\left(\frac{1-r}{1+r}\right)^{n/2}\right)\\
=&\frac{1}{r}
\sum_{k=0}^{[n/2]}
(k(k-1)+k)
\genfrac{(}{)}{0pt}{}{n}{k}
q(r)^k (1-q(r))^{n-k+1} \\
& -
\frac{1}{r}
\sum_{k=0}^{[n/2]-1}
(k(k-1)+ 3k +1)
\genfrac{(}{)}{0pt}{}{n}{k}
q(r)^{k+1} (1-q(r))^{n-k} \\
&+ O\left(\left(\frac{1-r}{1+r}\right)^{n/2}\right)\\
&\sum_{k=0}^{[n/2]}
k P_{n,r} (\frac{n}{2} -k) \\
=&
\sum_{k=0}^{[n/2]}
k \frac{1}{r}
\left(\genfrac{(}{)}{0pt}{}{n}{k}-
\genfrac{(}{)}{0pt}{}{n}{k-1}\right)
q(r)^k (1-q(r))^{n-k+1} \\
&+ O\left(\left(\frac{1-r}{1+r}\right)^{n/2}\right)\\
=&\frac{1}{r}
\sum_{k=0}^{[n/2]}
k 
\genfrac{(}{)}{0pt}{}{n}{k}
q(r)^k (1-q(r))^{n-k+1} \\
& -
\frac{1}{r}
\sum_{k=0}^{[n/2]-1}
(k+1) 
\genfrac{(}{)}{0pt}{}{n}{k}
q(r)^{k+1} (1-q(r))^{n-k} \\
&+ O\left(\left(\frac{1-r}{1+r}\right)^{n/2}\right).
\end{align*}
Furthermore, every term appearing in the above equation 
is calculated as
\begin{align*}
& \sum_{k=0}^{[n/2] \hbox{ or } [n/2]-1}
k
\genfrac{(}{)}{0pt}{}{n}{k}
q(r)^k (1-q(r))^{n-k} \\
=& 
\sum_{k=0}^{n}
k
\genfrac{(}{)}{0pt}{}{n}{k}
q(r)^k (1-q(r))^{n-k} \\
& +O ((1-r^2)^{n/2}) \\
=& n p +O ((1-r^2)^{n/2}) \\
&\sum_{k=0}^{[n/2]\hbox{ or } [n/2]-1}
k(k-1) 
\genfrac{(}{)}{0pt}{}{n}{k}
q(r)^k (1-q(r))^{n-k}\\
=&
\sum_{k=0}^{n}
k(k-1) 
\genfrac{(}{)}{0pt}{}{n}{k}
q(r)^k (1-q(r))^{n-k} \\
&+O ((1-r^2)^{n/2}) \\
= & n(n-1) p^2 
+O ((1-r^2)^{n/2}) .
\end{align*}
Note that
$(1-r^2) \,> \frac{1-r}{1+r}$.
Using there formulas, we obtain
\begin{align*}
& q(r)^2 +
\sum_{k=0}^{[n/2]}
P_{n,r} (n/2 -k)
(\frac{k^2}{n^2} - 2q(r)\frac{k}{n}) \\
=&
\frac{1-r^2}{4}\frac{1}{n}
- 
\frac{1-r}{2r }\frac{1}{n^2}
+ O ((1-r^2)^{n/2}),
\end{align*}
which implies (\ref{13-4}).
\endproof
\subsection{proof of (\ref{19-1})}\Label{19-2}
The left hand side of (\ref{19-1}) is calculated as
\begin{align}
& \sum_j 
\frac{1}{2^n}
\left(\genfrac{(}{)}{0pt}{}{n}{\frac{n}{2}-j}
-\genfrac{(}{)}{0pt}{}{n}{\frac{n}{2}-j-1}\right)
(2j+1) \left(\frac{2j}{n}\right)^2\nonumber \\
= &\sum_{k=1}^{[\frac{n}{2}]} 
\frac{1}{2^n}
\left(\genfrac{(}{)}{0pt}{}{n}{k}
-\genfrac{(}{)}{0pt}{}{n}{k-1}\right)
(n-2k+1) \left(\frac{n-2k}{n}\right)^2 \nonumber\\
&\quad +\frac{1}{2^n}\genfrac{(}{)}{0pt}{}{n}{0}
(n-2\cdot 0+1) \left(\frac{n- 2\cdot0}{n}\right)^2\nonumber\\
= &\frac{1}{n^2 2^n}
\Biggl(
\sum_{k=0}^{[\frac{n}{2}]} 
\genfrac{(}{)}{0pt}{}{n}{k}
(n-2k+1)(n-2k)^2 \nonumber\\
&\quad -\sum_{k=0}^{[\frac{n}{2}]-1} 
\genfrac{(}{)}{0pt}{}{n}{k}
(n-2k-1)(n-2k-2)^2
\Biggr) \Label{71-1}
\end{align}
When $n$ is even, $(n-2(\frac{n}{2})+1)(n-2(\frac{n}{2}))^2=0$ .
Then, the above value are calculated
\begin{align}
&\frac{1}{n^2 2^n}
\Biggl(
\sum_{k=0}^{[\frac{n}{2}]-1} 
\genfrac{(}{)}{0pt}{}{n}{k}
(n-2k+1)(n-2k)^2 \nonumber\\
&-\sum_{k=0}^{[\frac{n}{2}]-1} 
\genfrac{(}{)}{0pt}{}{n}{k}
(n-2k-1)(n-2k-2)^2
\Biggr) \nonumber\\
= &\frac{1}{n^2 2^n}
\left(
\sum_{k=0}^{[\frac{n}{2}]-1} 
\genfrac{(}{)}{0pt}{}{n}{k}
6(n-2k)^2 - 8 (n-2 k)+4
\right). \nonumber
\end{align}
The first term is calculated as
\begin{align*}
&\frac{1}{2^n}
\sum_{k=0}^{[\frac{n}{2}]-1} 
\genfrac{(}{)}{0pt}{}{n}{k}
6(n-2k)^2 \\
=&
\frac{1}{2^{n+1}}
\sum_{k=0}^{n} 
\genfrac{(}{)}{0pt}{}{n}{k}
6(2n (\frac{1}{2}- \frac{k}{n}))^2 - \genfrac{(}{)}{0pt}{}{n}{\frac{n}{2}}
6(n-2\frac{n}{2})^2 \\
= &
\frac{1}{2} \cdot 6 \cdot 4 n^2\cdot \frac{1}{4n}
= 3n .
\end{align*}
Since
$\sum_{k=0}^{[\frac{n}{2}]-1} \genfrac{(}{)}{0pt}{}{n-1}{k}
=\sum_{k=[\frac{n}{2}]}^{n-1} \genfrac{(}{)}{0pt}{}{n-1}{k}$,
we have
\begin{align*}
\frac{1}{2^n}
\sum_{k=0}^{[\frac{n}{2}]} 
\genfrac{(}{)}{0pt}{}{n}{k}
k =& 
\frac{n}{2^n}
\sum_{k=0}^{[\frac{n}{2}]-1} 
\genfrac{(}{)}{0pt}{}{n-1}{k} \\
= &
\frac{n}{2^{n+1}}
\sum_{k=0}^{n-1} 
\genfrac{(}{)}{0pt}{}{n-1}{k} 
= 
\frac{n}{4} \\
\frac{1}{2^n}
\sum_{k=0}^{[\frac{n}{2}]-1} 
\genfrac{(}{)}{0pt}{}{n}{k}
=&
\frac{1}{2^{n+1}}
\sum_{k=0}^{n}
\genfrac{(}{)}{0pt}{}{n}{k}
-
\genfrac{(}{)}{0pt}{}{n}{\frac{n}{2}}\frac{1}{2^{n+1}}\\
= &
\frac{1}{2} - \genfrac{(}{)}{0pt}{}{n}{\frac{n}{2}}\frac{1}{2^{n+1}} \\
\frac{1}{2^n}
\sum_{k=0}^{[\frac{n}{2}]} 
\genfrac{(}{)}{0pt}{}{n}{k}
=&\frac{1}{2} + \genfrac{(}{)}{0pt}{}{n}{\frac{n}{2}}\frac{1}{2^{n+1}} .
\end{align*}
Thus,
\begin{align*}
&\frac{1}{2^n}
\sum_{k=0}^{[\frac{n}{2}]-1} 
\genfrac{(}{)}{0pt}{}{n}{k}
\left(
- 8 (n- 2k)+4
\right) \\
= &\frac{1}{2^n}
\sum_{k=0}^{[\frac{n}{2}]} 
\genfrac{(}{)}{0pt}{}{n}{k}
\left(
- 8 (n- 2k)\right) 
4 \frac{1}{2} - \genfrac{(}{)}{0pt}{}{n}{\frac{n}{2}}\frac{1}{2^{n+1}} 
\\
= &- 8 ( \frac{n}{2} + 
n \genfrac{(}{)}{0pt}{}{n}{\frac{n}{2}}\frac{1}{2^{n+1}}
- \frac{n}{2})+ 
4 (\frac{1}{2} - \genfrac{(}{)}{0pt}{}{n}{\frac{n}{2}}\frac{1}{2^{n+1}}) \\
= & (-8 n -4)\genfrac{(}{)}{0pt}{}{n}{\frac{n}{2}}\frac{1}{2^{n+1}}) 
+ 2 .
\end{align*}
Since $
\genfrac{(}{)}{0pt}{}{n}{\frac{n}{2}}\frac{1}{2^{n+1}}
\cong \sqrt{\frac{1}{2\pi n}}$,
we have
\begin{align*}
\frac{3}{n}
-4 (2\frac{1}{n} +\frac{1}{n^2})
\genfrac{(}{)}{0pt}{}{n}{\frac{n}{2}}\frac{1}{2^{n+1}})
+ \frac{2 }{n^2} 
\cong
\frac{3}{n}
- \frac{4\sqrt{2}}{\sqrt{\pi}}\frac{1}{n \sqrt{n}}
+ \frac{2}{n^2}
\end{align*}

When $n$ is odd, $(n-2[\frac{n}{2}]-1)(n-2[\frac{n}{2}]-2)^2=0$ .
Then, the above value are calculated
\begin{align}
&\frac{1}{n^2 2^n}
\Biggl(
\sum_{k=0}^{[\frac{n}{2}]} 
\genfrac{(}{)}{0pt}{}{n}{k}
(n-2k+1)(n-2k)^2 \nonumber \\
&\quad -\sum_{k=0}^{[\frac{n}{2}]} 
\genfrac{(}{)}{0pt}{}{n}{k}
(n-2k-1)(n-2k-2)^2
\Biggr) \nonumber \\
= &\frac{1}{n^2 2^n}
\left(
\sum_{k=0}^{[\frac{n}{2}]} 
\genfrac{(}{)}{0pt}{}{n}{k}
6(n-2k)^2 - 8 (n-2 k)+4
\right) \Label{71-2}.
\end{align}
The first term is calculated as
\begin{align*}
\frac{1}{2^n}
\sum_{k=0}^{[\frac{n}{2}]} 
\genfrac{(}{)}{0pt}{}{n}{k}
6(n-2k)^2 
=&
\frac{1}{2^{n+1}}
\sum_{k=0}^{n} 
\genfrac{(}{)}{0pt}{}{n}{k}
6(2n (\frac{1}{2}- \frac{k}{n}))^2 \\
= &
\frac{1}{2} \cdot 6 \cdot 4 n^2\cdot \frac{1}{4n}
= 3n .
\end{align*}
Since
$\sum_{k=0}^{[\frac{n}{2}]-1} \genfrac{(}{)}{0pt}{}{n-1}{k}
=\sum_{k=[\frac{n}{2}]+1}^{n-1} \genfrac{(}{)}{0pt}{}{n-1}{k}$,
we have
\begin{align*}
& \frac{1}{2^n}
\sum_{k=0}^{[\frac{n}{2}]} 
\genfrac{(}{)}{0pt}{}{n}{k}
k = 
\frac{n}{2^n}
\sum_{k=0}^{[\frac{n}{2}]-1} 
\genfrac{(}{)}{0pt}{}{n-1}{k} \\
=&
\frac{n}{2^{n+1}}
\sum_{k=0}^{n-1} 
\genfrac{(}{)}{0pt}{}{n-1}{k}
-\frac{n}{2^{n+1}} \genfrac{(}{)}{0pt}{}{n-1}{[\frac{n}{2}]} \\
= &
\frac{n}{4}
-\frac{n}{2^{n+1}} \genfrac{(}{)}{0pt}{}{n-1}{[\frac{n}{2}]},
\end{align*}
and 
\begin{align*}
& \frac{1}{2^n}
\sum_{k=0}^{[\frac{n}{2}]} 
\genfrac{(}{)}{0pt}{}{n}{k}
=
\frac{1}{2^{n+1}}
\sum_{k=0}^{n}
\genfrac{(}{)}{0pt}{}{n}{k}
= \frac{1}{2} .
\end{align*}
Since $
\genfrac{(}{)}{0pt}{}{n}{[\frac{n}{2}]}\frac{1}{2^{n}}
\cong \sqrt{\frac{1}{\pi [\frac{n}{2}]}}$,
(\ref{71-2}) can be approximated as
\begin{align*}
&{\rm R.H.S. of (\ref{71-2})} \\
= &\frac{1}{n^2}\left(
3n +(- 8n+ 4) \frac{1}{2}+ 16 
\left(
\frac{n}{4}
-\frac{n}{2^{n+1}} \genfrac{(}{)}{0pt}{}{n-1}{[\frac{n}{2}]}
\right)
\right)\\
=& \frac{1}{n^2}\left(
3n + 2 - 16 
 \frac{n}{2^{n}} \genfrac{(}{)}{0pt}{}{n-1}{[\frac{n}{2}]} \right)\\
\cong &
\frac{3}{n} - \frac{8}{ n \sqrt{\pi [\frac{n}{2}]}}+ \frac{2}{n^2} 
\cong 
\frac{3}{n} - \frac{4 \sqrt{2}}{ \sqrt{\pi }n \sqrt{n}}+ \frac{2}{n^2} .
\end{align*}
\endproof
\subsection{Proof of Lemma \ref{le27-14}}\Label{le27-14-pf}
First, we parameterize the square root of $G$ as
\begin{align*}
\sqrt{G}= \left(
\begin{array}{cc}
A & a \\
a^T & t
\end{array}
\right),
\end{align*}
where $A$ is a $2\times 2$ symmetric 
matrix and $a$ is a 2-dimensional vector.
\begin{align*}
&C_{(0,0,r)}^R(G)\\
=&
\tr G -r^2 s^2
+ 
r
\tr \left| 
\left(
\begin{array}{cc}
A & a \\
a^T & 0 
\end{array}
\right)
\left(
\begin{array}{ccc}
0 & -i & 0 \\
i & 0  & 0 \\
0 & 0  & 0
\end{array}
\right)
\left(
\begin{array}{cc}
A & a \\
a^T & 0 
\end{array}
\right)
\right|.
\end{align*}
By putting $J= \left(
\begin{array}{cc}
0 & -1 \\
1 & 0
\end{array}
\right)$, we can calculated the second term as:
\begin{align*}
&\tr \left| 
\left(
\begin{array}{cc}
A & a \\
a^T & 0 
\end{array}
\right)
\left(
\begin{array}{ccc}
0 & -i & 0 \\
i & 0  & 0 \\
0 & 0  & 0
\end{array}
\right)
\left(
\begin{array}{cc}
A & a \\
a^T & 0 
\end{array}
\right)
\right|\\
= &
\tr \left|
i
\left(
\begin{array}{cc}
(\det A) J & A J a\\
(A J a)^T & 0\\
\end{array}
\right)
\right|\\
= &
2 \sqrt{(\det A)^2 + \|A J a\|^2}
=
2 \sqrt{(\det A^2) + \langle J a |A^2 |J a\rangle}\\
=&
2 \sqrt{\det (A^2 +  | a \rangle \langle a|)},
\end{align*}
where 
the final equation 
can be checked by choosing
a basis such that $
a= \left(
\begin{array}{c}
\|a\| \\
0
\end{array}
\right)$.
Since $\tilde{G}=
A^2 +  | a \rangle \langle a|$,
we obtain (\ref{15-1}).
\endproof

\subsection{Proof of (\ref{10-3-7})}\Label{10-3-6}
First, we focus the following expressions of 
$C_{\theta}^R(G)$ and $\hat{C}_{\theta}(G)$
\begin{align}
\hat{C}_{\theta}(G)& 
= \left( \tr \sqrt{\sqrt{G}J_\theta^{-1}\sqrt{G}}\right)^2 \\
C_{\theta}^R(G)& = 
\tr \sqrt{G}J_\theta^{-1}\sqrt{G}+ 
\tr |  2 \sqrt{G}J_\theta^{-1} D_\theta J_\theta^{-1} \sqrt{G}|.
\end{align}
When we put 
the real symmetric matrix $A \defeq \sqrt{G}J_\theta^{-1}\sqrt{G}$ 
and the real anti-symmetric matrix 
$B \defeq 2 \sqrt{G}J_\theta^{-1} D_\theta J_\theta^{-1} \sqrt{G}$,
the relation 
\begin{align}
A + i B \ge 0.
\end{align}
Here, we dragonize $A$ as
\begin{align}
A= \left(
\begin{array}{ccc}
a & 0 & 0 \\
0 & b & 0 \\
0 & 0 & c
\end{array}
\right)
\end{align}
with $a \ge b \ge c$ and $c > 0$,
where the final strict inequality 
follows from $G \,> 0$.
Since $|B|$ is a constat times of a two-dimensional projection $P$.
Hence,
\begin{align}
P A P + iB = P (A + iB)P \ge 0.
\end{align}
If we regard $PAP$ as a two-dimensional matrix,
$\tr |B| \le 2 \sqrt{\det PAP}$.
Thus, by considering the maximum case of the minimum eigen value of 
$PAP$, we have
\begin{align}
\tr |B| \le 2 \sqrt{a b}
\end{align}
Therefore,
\begin{align*}
&\hat{C}_{\theta}(G)- C_{\theta}^R(G)
= (\tr \sqrt{A})^2 - (\tr A + \tr |B|) \\
\ge &
2\left(\sqrt{ab} + \sqrt{bc} + \sqrt{ca}
-  \sqrt{a b}
\right)
= 2\left(\sqrt{bc} + \sqrt{ca}
\right)\,> 0.
\end{align*}
\endproof
\subsection{Proofs of (\ref{17-4-1}) and (\ref{17-4-2})}
\Label{27-16}
Since 
\begin{align*}
\int_{\real} \hat\theta^k
V_j^* M_{\tilde{G}}(\,d \hat\theta) V_j 
= \left\{
\begin{array}{ll}
Q& k=1 \\
P& k=2
\end{array}
\right. ,
\end{align*}
we have
\begin{align*}
\Tr (\rho_{j,p} \circ 2 J_{j,k}) 
\int_{\real} 
\hat\theta^l
M_{j,\tilde{G}}(\,d \hat\theta)
=\delta_k^l 
\end{align*}
for $k,l = 1,2 $, where 
\begin{align*}
M_{j,\tilde{G}}(\,d \hat\theta)
\defeq
V_j^* M_{\tilde{G}}\circ B_j^{-1}(\,d \hat\theta) V_j .
\end{align*}
Since 
the matrixes $\rho_{j,p} \circ \frac{1}{1-r^2}(2 J_{j,3} - r I) $
and
$\rho_{j,p}$
are diagonal
and all diagonal elements of 
$V_j^* Q V_j $ and $V_j^* P V_j$ are $0$,
we have
\begin{align*}
\Tr (\rho_{j,p} \circ \frac{1}{1-r^2}(2 J_{j,3} - r I) )
V_j^* Q V_j &= 
\Tr \rho_{j,p} V_j^* Q V_j = 0 \\
\Tr (\rho_{j,p} \circ \frac{1}{1-r^2}(2 J_{j,3} - r I) )
V_j^* P V_j &= 
\Tr \rho_{j,p} V_j^* P V_j = 0 .
\end{align*}
Thus,
\begin{align}
& \Tr (\rho_{j,p} \circ \frac{1}{1-r^2}(2 J_{j,3} - r I) ) 
\int_{\real} 
\hat\theta^l
M_{j,\tilde{G}}(\,d \hat\theta) \nonumber\\
= &
\Tr \rho_{j,p} 
\int_{\real} 
\hat\theta^l
M_{j,\tilde{G}}(\,d \hat\theta)
=0 \Label{16-8}
\end{align}
for $l=1,2$.
Therefore,
\begin{align*}
&\Tr \frac{\partial \rho_\theta^{\otimes n}}
{\partial \theta^k}
\left(\bigoplus_j 
\left(\int_{\real} 
\hat\theta^l
M_{j,\tilde{G}}(\,d \hat\theta)\right)
\otimes I_{{\cal H}_{n,j}}\right) \nonumber \\
= &
\sum_{j} P_{n,r}(j)
\Tr (\rho_{j,p} \circ 2 J_{j,k}) 
\int_{\real} 
\hat\theta^l
M_{j,\tilde{G}}(\,d \hat\theta) 
= \delta_k^l 
\end{align*}
for $k,l= 1,2$.
For the $k=3$ case,
the above quantity equals
\begin{align*}
& \sum_{j} P_{n,r}(j)
\Tr (\rho_{j,p} \circ \frac{1}{1-r^2}(2 J_{j,3} - r I) ) 
\int_{\real} 
\hat\theta^l
M_{j,\tilde{G}}(\,d \hat\theta) =0 .
\end{align*}
Furthermore,
we have 
\begin{align}
&\Tr \rho_{(0,0,r)}^{\otimes n}
\left(\bigoplus_j 
\left(\int_{\real} 
\hat\theta^l
M_{j,\tilde{G}}(\,d \hat\theta)\right)
\otimes I_{{\cal H}_{n,j}}\right) \nonumber \\
=&
\sum_{j} P_{n,r}(j)
\Tr \rho_{j,p} 
\int_{\real} 
\hat\theta^l
M_{j,\tilde{G}}(\,d \hat\theta)
=0
\Label{17-6}
\end{align}
for $l=1,2$.
\endproof
\subsection{Proof of (\ref{19-3})}\Label{19-4}
First, we focus on the following equation
\begin{align*}
& \frac{n}{1-r^2} \\
= &
\Tr \rho_{(0,0,r)}^{\otimes n}
\left(\bigoplus_j
\frac{1}{1-r^2}( 2 J_{j,3} -r I) \otimes I_{{\cal H}_{n,j}}
\right)^2\\
=&\sum_j
P_{n,r}(j)
\sum_{m=-j}^j 
\frac{1-p}{1-p^{2j+1}}p^{j-m} 
\left(\frac{1}{1-r^2}\right)^2
(2 m - r)^2 .
\end{align*}
Then, applying Theorem \ref{29-1-1},
we can see that the difference $\frac{n}{1-r^2}- J_{n,r}$ equals
information loss. Thus,
\begin{align*}
\frac{n}{1-r^2}
- J_{n,r} = \sum_j P_{n,r}(j) \tilde{J}_{j,r},  
\end{align*}
where
\begin{align*}
\tilde{J}_{j,r} \defeq &
\sum_{m=-j}^j 
\frac{1-p(r)}{1-p(r)^{2j+1}}p(r)^{j-m} 
\left(\frac{2m-r}{1-r^2}\right)^2  \\
& - 
\left(\sum_{m=-j}^j 
\frac{1-p(r)}{1-p(r)^{2j+1}}p(r)^{j-m} \frac{2m-r}{1-r^2}\right)^2
\\
= &
\sum_{m=-j}^j 
\frac{1-p(r)}{1-p(r)^{2j+1}}p(r)^{j-m} 
\left(\frac{2(m-j)}{1-r^2}\right)^2  \\
&- 
\left(\sum_{m=-j}^j 
\frac{1-p(r)}{1-p(r)^{2j+1}}p(r)^{j-m} \frac{2(m-j)}{1-r^2}
\right)^2 \\
= &
\frac{4(1-p(r))}{(1-p(r)^{2j+1})(1-r^2)^2} 
\Biggl(
\frac{p(r)+p(r)^2}{(1-p(r))^3} \\
& -\frac{(1+p(r))p(r)^{2j+1}}{(1-p(r))^3} \\
& -\frac{(4j + (2j)^2(1-p(r)))p(r)^{2j+1}}{(1-p(r))^2}
\Biggr) \\
&-
\frac{4(1-p(r))^2}{(1-p(r)^{2j+1})^2(1-r^2)^2} \\
& \quad \times \left(
\frac{p(r)(1-p(r)^{2j})}{(1-p(r))^2} 
- \frac{2j p(r)^{2j+1}}{1-p(r)}
\right)^2 \\
= &
\frac{1}{r^2(1-r^2)} + O(p(r)^{2j}).
\end{align*}
Thus,
\begin{align*}
\frac{n}{1-r^2}- J_{n,r}\to \frac{1}{r^2(1-r^2)},
\end{align*}
which implies (\ref{19-3}).
\endproof

\begin{widetext}
\subsection{Proof of (\ref{19-5})}\Label{19-6}
For the covariance of the POVM $M_{{\rm cov}}^n$,
the Fisher information matrix $J_{(0,0,0)}^{M_{{\rm cov}}^n}$
is a scalar times of the identical matrix.
We apply Theorem \ref{29-1} to the family of probability 
distributions 
$p_r(j,\phi,\psi)\defeq
\Tr \rho_{(0,0,r)}^{\otimes n} 
M^j(\phi,\psi) \otimes I_{{\cal H}_{n,j}}
= P_{n,r}(j) \Tr \rho_{j,p} M^j (\phi,\psi)$.
Then, we calculate the Fisher information:
\begin{align*}
J_{{\rm cov}}^n 
= 
\sum_j 
P_{n,r}(j)
\left(\frac{\,d P_{n,r}(j) }{\,d r}\right)^2 
 +
\sum_j 
P_{n,r}(j)
\int 
\left(\frac{\,d \Tr \rho_{j,p} M^j (\phi,\psi)}{\,d r}
\right)^2 
\Tr \rho_{j,p} M^j (\phi,\psi)
\,d \phi \,d \psi.
\end{align*}
On the other hand, 
Applying Theorem \ref{29-1-1} to
$P_{n,r}(j) \langle j,m| \rho_{j,p}|j,m\rangle$,
we have 
\begin{align*}
 \frac{n}{1-r^2} \\
=
\sum_j 
\left(\frac{\,d P_{n,r}(j)}{\,d r}\right)^2
P_{n,r}(j)
+
\sum_j 
P_{n,r}(j)
\sum_{m=-j}^j 
\left(\frac{\,d \langle j,m| \rho_{j,p}|j,m \rangle}{\,d r}\right)^2
\langle j,m| \rho_{j,p}|j,m\rangle.
\end{align*}
Thus,
\begin{align*}
J_{{\rm cov}}^n
= &\frac{n}{1-r^2} 
- 
\sum_j 
P_{n,r}(j)
\Biggl(\int 
\left(\frac{\,d \Tr \rho_{j,p} M^j (\phi,\psi)}{\,d r}
\right)^2
\Tr \rho_{j,p} M^j (\phi,\psi)
\,d \phi \,d \psi \\
& -\sum_{m=-j}^j 
\left(\frac{\,d  \langle j,m| \rho_{j,p}|j,m \rangle}{\,d r}\right)^2
\langle j,m| \rho_{j,p}|j,m \rangle \Biggr).
\end{align*}
In the case of $r=0$, Since
$\Tr \rho_{j,p} M^j (\phi,\psi)
= 
(2j+1)\frac{1-p(r)}{1-p(r)^{2j+1}}
\left(\frac{1+r \cos \phi}{1+r}\right)^{2j} \frac{\sin \phi}{4\pi}
$, we obtain
\begin{align*}
&\int_0^{2\pi} \int_0^{\pi} 
\left(\frac{\,d \log \Tr \rho_{j,p(r)} M^j (\phi,\psi)}{\,d r}
\right)^2
\Tr \rho_{j,p(r)} M^j (\phi,\psi)
\,d \phi \,d \psi \\
=&
\int_0^{2\pi} \int_0^{\pi} 
\left(\frac{\,d \log \frac{1-p(r)}{1-p(r)^{2j+1}}
\left(\frac{1+r \cos \phi}{1+r}\right)^{2j} }{\,d r}\right)^2
(2j+1)\frac{1-p(r)}{1-p(r)^{2j+1}}
\left(\frac{1+r \cos \phi}{1+r}\right)^{2j} \frac{\sin \phi}{4\pi}
\,d \phi \,d \psi  \\
=&
\int_0^{2\pi} \int_0^{\pi} 
\left(
\frac{\,d \log 
     \left(\frac{1+r \cos \phi}{1+r}\right)^{2j} 
     }{\,d r}
\right)^2
(2j+1)\frac{1-p(r)}{1-p(r)^{2j+1}}
\left(\frac{1+r \cos \phi}{1+r}\right)^{2j} 
\frac{\sin \phi}{4\pi}
\,d \phi \,d \psi  
+
\left(\frac{\,d \log \frac{1-p(r)}{1-p(r)^{2j+1}}}
{\,d r}\right)^2.
\end{align*}
Its first and second terms are calculated as
\begin{align*}
&\int_0^{2\pi} \int_0^{\pi} 
\left(
\frac{\,d \log 
\left(\frac{1+r \cos \phi}{1+r}\right)^{2j} }{\,d r}\right)^2
(2j+1)\frac{1-p(r)}{1-p(r)^{2j+1}}
\left(\frac{1+r \cos \phi}{1+r}\right)^{2j} \frac{\sin \phi}{4\pi}
\,d \phi \,d \psi  
= \int_0^\pi 
\frac{1}{2} (2j \cos \phi )^2\sin \phi \,d \phi =
\frac{4}{3}j^2 ,\\
& \frac{\,d \log \frac{1-p(r)}{1-p(r)^{2j+1}}}{\,d r}
=- \int_0^{2\pi} \int_0^{\pi} 
\frac{\,d \log 
     \left(\frac{1+r \cos \phi}{1+r}\right)^{2j} 
     }{\,d r}
(2j+1)\frac{1-p(r)}{1-p(r)^{2j+1}}
\left(\frac{1+r \cos \phi}{1+r}\right)^{2j} \frac{\sin \phi}{4\pi}
\,d \phi \,d \psi  \\
= &\int_0^\pi 
\frac{1}{2} 2j \cos \phi \sin \phi \,d \phi =0 .
\end{align*}
\end{widetext}
Since
\begin{align*}
-\sum_{m=-j}^j 
\left(\frac{\,d \langle j,m| \rho_{j,p}|j,m\rangle}{\,d r}\right)^2
\langle j,m| \rho_{j,p}|j,m \rangle 
= \frac{4}{3}j(j+1),
\end{align*}
we have
\begin{align*}
J_{{\rm cov}}^n
= n- \sum_j 
P_{n,0}(j) \frac{4}{3}j  
\end{align*}
Therefore, if the relation
\begin{align}
P_{n,0}(j) \frac{4}{3}j  \cong & 
\frac{4 \sqrt{2}}{3\sqrt{\pi} }\sqrt{n} +\frac{2}{3} \Label{71-7}
\end{align}
holds, we obtain (\ref{19-5}).
Hence, in the following, we will prove (\ref{71-7}).
\begin{align}
&\sum_j 
P_{n,0}(j) \frac{4}{3}j \nonumber \\
=& \frac{2}{3}
\sum_j 
\frac{1}{2^n}
\left(\genfrac{(}{)}{0pt}{}{n}{\frac{n}{2}-j}
-\genfrac{(}{)}{0pt}{}{n}{\frac{n}{2}-j-1}\right)
2j(2j+1)\nonumber \\
= &\frac{2}{3} \sum_{k=1}^{[\frac{n}{2}]} 
\frac{1}{2^n}
\left(\genfrac{(}{)}{0pt}{}{n}{k}
-\genfrac{(}{)}{0pt}{}{n}{k-1}\right)
(n-2k+1)(n-2k)\nonumber\\
& +\frac{2}{3}\frac{1}{2^n}\genfrac{(}{)}{0pt}{}{n}{0}
(n-2\cdot 0+1)(n-2\cdot 0)
\nonumber\\
= &\frac{2}{3}\frac{1}{2^n}
\Biggl(
\sum_{k=0}^{[\frac{n}{2}]} 
\genfrac{(}{)}{0pt}{}{n}{k}
(n-2k+1)(n-2k) \nonumber\\
&-\sum_{k=0}^{[\frac{n}{2}]-1} 
\genfrac{(}{)}{0pt}{}{n}{k}
(n-2k-1)(n-2k-2)
\Biggr) \Label{71-5}.
\end{align}
When $n$ is even, $(n-2(\frac{n}{2})+1)(n-2(\frac{n}{2}))^2=0$ .
Then, the above value are calculated
\begin{align*}
&\sum_j 
P_{n,0}(j) \frac{4}{3}j \\
=
&\frac{2}{3 2^n}
\Biggl(
\sum_{k=0}^{[\frac{n}{2}]-1} 
\genfrac{(}{)}{0pt}{}{n}{k}
(n-2k+1)(n-2k)\\
&-\sum_{k=0}^{[\frac{n}{2}]-1} 
\genfrac{(}{)}{0pt}{}{n}{k}
(n-2k-1)(n-2k-2)
\Biggr) \\
= &\frac{1}{3 2^n}
\left(
\sum_{k=0}^{[\frac{n}{2}]-1} 
\genfrac{(}{)}{0pt}{}{n}{k}
8 (n-2 k)-4
\right) 
\cong \frac{4 \sqrt{2}}{3\sqrt{\pi} }\sqrt{n} +\frac{2}{3} .
\end{align*}
When $n$ is odd, $(n-2[\frac{n}{2}]-1)(n-2[\frac{n}{2}]-2)^2=0$ .
Then, the above value are calculated
\begin{align*}
&\sum_j 
P_{n,0}(j) \frac{4}{3}j \\
=&\frac{2}{3 2^n}
\Biggl(
\sum_{k=0}^{[\frac{n}{2}]} 
\genfrac{(}{)}{0pt}{}{n}{k}
(n-2k+1)(n-2k) \\
&-\sum_{k=0}^{[\frac{n}{2}]} 
\genfrac{(}{)}{0pt}{}{n}{k}
(n-2k-1)(n-2k-2)
\Biggr) \\
= &\frac{1}{3 2^n}
\left(
\sum_{k=0}^{[\frac{n}{2}]} 
\genfrac{(}{)}{0pt}{}{n}{k}
8 (n-2 k)-4
\right) 
 \cong \frac{4 \sqrt{2}}{3\sqrt{\pi} }\sqrt{n} +\frac{2}{3} ,
\end{align*}
which implies (\ref{71-7}).
\endproof
\subsection{Proof of Theorem \ref{27-18}}\Label{27-18-1}
We focus on the full parameter model with the derivatives
at the point $(r,0,0)$
\begin{align}
\frac{\partial \rho}{\partial \theta^1}=
\sigma_1, 
\frac{\partial \rho}{\partial \theta^2}=
\sigma_2, 
\frac{\partial \rho}{\partial \theta^3}=
(1-r^2)\sigma_3
\end{align}
as a D-invariant model.
In this case, the SLD Fisher information matrix is the identity matrix.
Thus, we can apply (\ref{10-19-1}) of Lemma \ref{le-3}.
Hence,
by putting
\begin{align*}
d_1=\left(
\begin{array}{c}
1 \\ 
0 \\
0
\end{array}
\right),\quad
d_2=\left(
\begin{array}{c}
0 \\ 
\cos \phi \\
\sin \phi
\end{array}
\right),
\end{align*}
we obtain 
\begin{align*}
C^H_\theta(G)
= \min_{v=[v^j]}\left\{\left.
\tr | \sqrt{G} 
Z_J(v)
\sqrt{G} |
\right| \re \langle d_k|J|v^j\rangle = \delta_k^j\right\},
\end{align*}
where
\begin{align*}
J\defeq
\left(
\begin{array}{ccc}
1 & -ir & 0 \\
ir & 1 & 0 \\
0 & 0 & 1 
\end{array}
\right).
\end{align*}
Hence, from the condition
\begin{align*}
\langle d_j|J| v^k\rangle = \delta_j^k.
\end{align*}
Then, $v^1$ and $v^2$ are parameterized as
\begin{align*}
v^1  = & L_1 - t \sin \phi L_2 
 + t \cos \phi L_3 \\
v^2  = & (- s \sin \phi + \cos \phi)L_2 
+ (s \cos \phi + \sin \phi)L_3 .
\end{align*}
The matrix $Z_J(v)$ can be calculated as
\begin{align*}
\left(
\begin{array}{cc}
1 + t^2 & t s - ir ( -s \sin \phi + \cos \phi) \\
t s + ir ( -s \sin \phi + \cos \phi) & 1+ s^2
\end{array}
\right).
\end{align*}
Thus, the quantity $\tr |\sqrt{G} Z_J(v)\sqrt{G}|$ equals
\begin{align}
\tr G + g_1 (t + \frac{g_2}{g_1}s)^2 +
\frac{\det G}{g_1} s^2 + 
2r | \cos \phi- \sin \phi s| 
\sqrt{\det G} \Label{23-2}.
\end{align}
In the following, we treat the case of 
$\frac{g_1}{\sqrt{\det G}}\,< \frac{\cos \phi}{r \sin^2 \phi}$.
The minimum value of (\ref{23-2})
equals
$\tr G + 2r \cos \phi \det G - r^2 \sin^2 \phi g_1$
which is attained by the parameters
$t= - \frac{g_2}{g_1}s, s= \frac{r g_1 \sin \phi }{\sqrt{\det G}}$.
Thus, 
the discussion in subsubsection \ref{9-29-8} guarantees that 
the Holevo bound is attained only by the following covariance matrix 
\begin{align}
& \re Z_\theta(\vec{X}) +
\sqrt{G}^{-1} | \sqrt{G} \im Z_\theta(\vec{X})\sqrt{G}|\sqrt{G}^{-1} 
\nonumber \\
= &
\left(
\begin{array}{cc}
1 + t^2 & t s  \\
t s & 1+ s^2
\end{array}
\right)
+ r | -s \sin \phi + \cos \phi|
\sqrt{\det G}G^{-1} \Label{23-8}\\
= & \hbox{R.H.S. of } (\ref{23-6}).\nonumber 
\end{align}
In the opposite case,
the minimum value of (\ref{23-2})
equals
R.H.S. of (\ref{23-7}),
which is attained by the parameters
$t= - \frac{g_2}{g_1}s, s= \frac{\cos \phi }{\sin \phi}$.
Substituting these parameters into (\ref{23-8}),
we obtain (\ref{23-9}).
\endproof

\end{document}